\documentclass[preprint, superscriptaddress, nofootinbib]{revtex4-1}
\usepackage{graphicx} 
\usepackage[colorlinks=true,linkcolor=blue,citecolor=blue,urlcolor=black]{hyperref}
\usepackage[english]{babel}
\usepackage{datetime}
\usepackage{verbatim}
\usepackage{amsmath}
\usepackage{amssymb}
\usepackage{siunitx} 
\usepackage{nccmath}
\usepackage{dcolumn}
\usepackage{lineno}
\usepackage{xcolor}
\usepackage{color}
\usepackage{float}
\usepackage[utf8]{inputenc}
\usepackage{lineno}
\usepackage[symbolgreek]{mathastext}
\usepackage{import}
\usepackage{marvosym}

\begin{document}

\widetext
\raggedbottom

\title{\Large Gating and tunable confinement of active colloids within patterned environments}

\author{Carolina van Baalen$^{~\dagger}$ }
\affiliation{Laboratory for Soft Materials and Interfaces, Department of Materials, ETH Zürich, Switzerland. \\E-mail: lucio.isa@mat.ethz.ch}

\author{Stefania Ketzetzi$^{~\dagger}$}
\affiliation{Laboratory for Soft Materials and Interfaces, Department of Materials, ETH Zürich, Switzerland. \\E-mail: lucio.isa@mat.ethz.ch}
\affiliation{Present address: John A. Paulson School of Engineering and Applied Sciences, Harvard University, Cambridge, MA 02138, USA.}

\author{Anushka Tintor}
\affiliation{Laboratory for Soft Materials and Interfaces, Department of Materials, ETH Zürich, Switzerland. \\E-mail: lucio.isa@mat.ethz.ch}

\author{Lucio Isa}
\affiliation{Laboratory for Soft Materials and Interfaces, Department of Materials, ETH Zürich, Switzerland. \\E-mail: lucio.isa@mat.ethz.ch}

\let\thefootnote\relax\footnotetext{\textit{$^{\dagger}$~These authors contributed equally to this work.}}

\date{\today}

\begin{abstract}

Active colloidal particles typically exhibit a pronounced affinity for accumulating and being captured at boundaries. Here, we engineer long-range repulsive interactions between colloids that self-propel under an electric field and patterned obstacles. As a result of these interactions, particles turn away from obstacles and avoid accumulation. We show that by tuning the applied field frequency, we precisely and rapidly control the effective size of the obstacles and therefore modulate the particle approach distance. This feature allows us to achieve gating and tunable confinement of our active particles whereby they can access regions between obstacles depending on the applied field. Our work provides a versatile means to directly control confinement and organization, paving the way towards applications such as sorting particles based on motility or localizing active particles on demand.

\end{abstract}

\maketitle

\subsection*{\Large Introduction}

Biological microswimmers employ sensory and feedback control schemes to enable a plethora of interactions with confining surfaces based on decision making and active response~\cite{Moreno2017, Lele2013, Persat2015, Laventie2020, Singh2021, Lin2015, Stoodley2004, Lee2020, DiLuzio2005, Drescher2011, Kantsler2013, Berne2016}. Information exchange with their surroundings leads to targeted motility modes and behaviors, including alignment, capture, and accumulation, \textit{e.g.}, when colonizing surfaces, or avoidance and escape at boundaries, \textit{e.g.}, when exploring space to ensure proper dispersion~\cite{Lauga2009, Elgeti2015}. In recent decades, scientists have been developing active colloidal particle systems to serve as synthetic model microswimmers~\cite{Dreyfus2005, Golestanian2005, Bechinger2016, Zottl2016}. Owing to their ability to convert energy from the environment into directed motion, these are outstanding candidates for technological and industrial applications~\cite{Katuri2017, Patra2013, Han2018,Garcia2013, Restrepo2014, Gao2015, Wang2018, Ceylan2019,Gao2013, Wang2021, Xu2021}. However, the vision that they will surpass the range of functions and applicability of the biological systems that inspire them still remains unattainable to date~\cite{Ebbens2016, Bishop2023, Tsang2020}.

Contrary to their biological counterparts, active colloids still lack targeted motility modes and comparable behaviors in relation to confining boundaries. Instead, they exhibit ``involuntary" affinity and thus pronounced accumulation at surfaces~\cite{Mijalkov2013, Katuri2018, Maggi2015,Reichhardt2017, Kaiser2012, Palacios2021}. First, active systems, \textit{e.g.}, the traditional Janus particles that catalytically self-propel in H$_2$O$_2$ solutions, are bound to move along the walls of their container, due to the competition between gravity and mass asymmetry on their bodies~\cite{Ebbens2011, Campbell2013, Ketzetzi2020sep, Bailey2024, Carrasco2023}. In addition, catalytic microswimmers are captured by secondary structures fixed on the walls due to hydrodynamic and phoretic couplings~\cite{Das2015,Simmchen2016,Yu2016,Liu2016}. Said structures act as obstacles causing permanent immobilization, or alignment followed by sliding and orbiting~\cite{Takagi2014, Brown2016, Baalen2023, Ketzetzi2022}. Overall, escape from obstacles is rare and poorly controlled, as it stems from variations inherent to the static shape of the confinement~\cite{Ketzetzi2022, Simmchen2016, Wykes2017}. Designing tunable interactions between active colloids and obstacles may therefore enable novel behavior and modes of active transport across environments on both the single-particle and the collective level, \textit{e.g.}, allowing particles to follow complex paths and realizing environments for gating and tunable confinement.

Here, we engineer long-range repulsive microswimmer-boundary interactions inside patterned environments. We employ synthetic microswimmers in the form of metallo-dielectric colloids self-propelling under an alternating current (AC) electric field. Unlike catalytically-active colloids, our colloids approach and turn away from obstacles with remarkable robustness, without accumulation or self-trapping~\cite{Tanuku2023}. We demonstrate that this turning-away response can be directly adapted via the field frequency, which sets a repulsive exclusion zone around the obstacle. This effectively tunes the obstacle size, turning each obstacle into a ``soft" repulsive boundary. By tuning the frequency, we thus dynamically shape the boundaries experienced by the active colloids to direct them along straight paths or around bends, and to achieve gating and confinement as a function of effective distances between obstacles. Overall, our work provides an exciting and versatile way to direct synthetic microswimmer motion and organization within complex environments, which can prove useful for future applications towards sorting particles based on their motility or creating photonic devices with active particles~\cite{Trivedi2022}.

\begin{figure*}[!ht]
    \centering
    \includegraphics[width=1\linewidth]{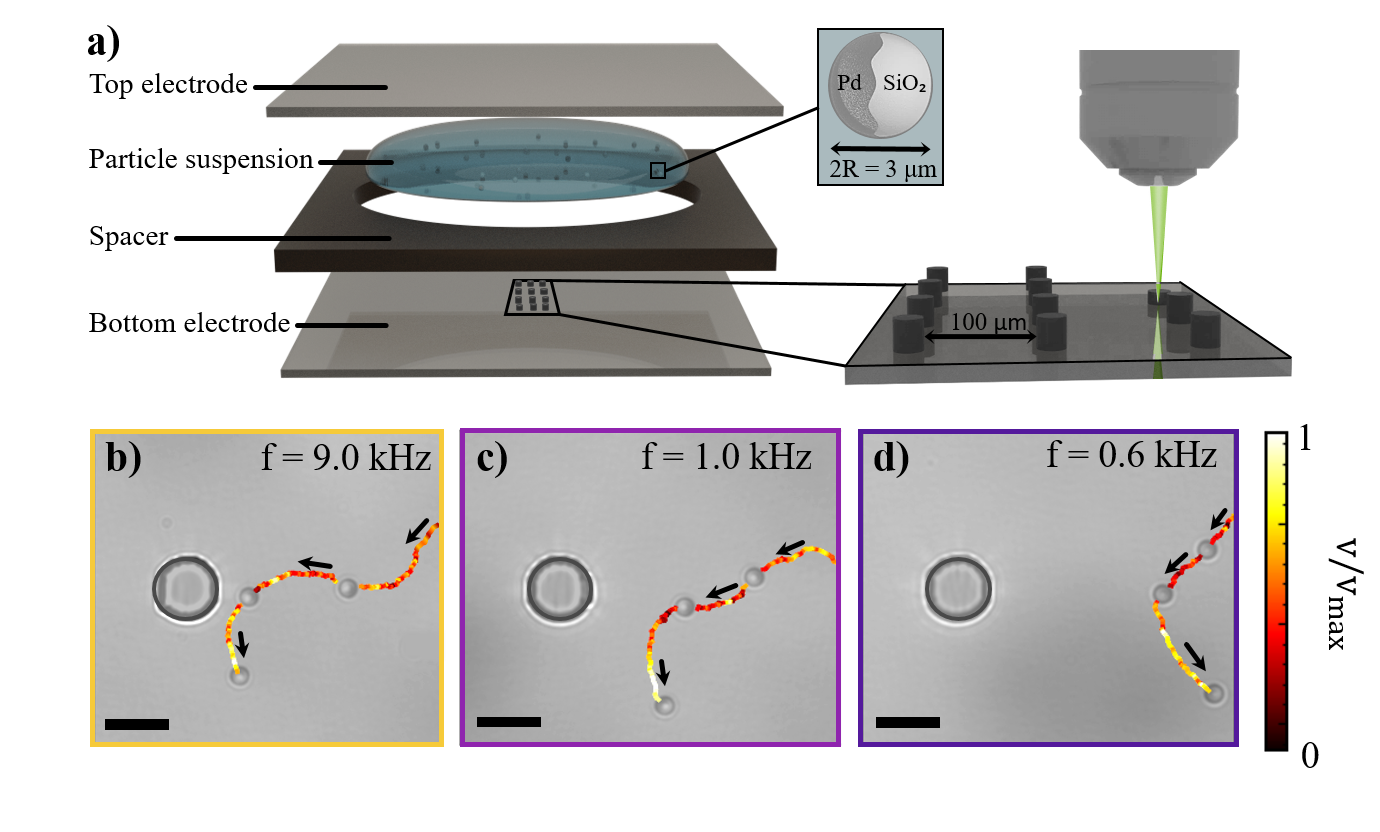}\vspace{-25pt}
    \caption{\textbf{Synthetic microswimmers turn away from obstacles.} a) Our experimental system: we study the active motion of 3 $\mu$m SiO$_2$ Janus colloids with thin Pd caps self-propelling under an AC electric field. We place the colloids in water between two transparent conductive surfaces separated by a spacer and connected to a function generator, and image particle motion with an inverted microscope. The bottom surface, parallel to which colloids self-propel, is patterned with 3D-printed obstacles in various configurations. Upon approach, the active particles ``sense" and avoid the obstacle surface due to tunable long-range repulsive interactions. Microscopy images obtained at a fixed 6 V peak-to-peak amplitude and frequency b) 9.0 kHz, c) 1.0 kHz, and d) 0.6 kHz, illustrating the turning-away response. Trajectories are colored according to the normalized instantaneous velocity of the particle. Scale bars are 10 $\mu$m. 
    }
    \label{fig:Fig1}
\end{figure*}

\subsection*{\Large Results and Discussion} 
\subsection*{\large Tunable long-ranged interactions between microswimmers and obstacles instigate novel turn away response}

In our experiments, we explore the interaction between Janus colloids self-propelling under an AC electric field~\cite{Shields2017} and static 3D-printed obstacles. To that end, we employ metallo-dielectric spheres comprising a 3 $\mu$m diameter SiO$_2$ core half-coated with a thin layer of Pd, see Appendix I. We place the Janus colloids in water between two transparent conductive surfaces separated by a spacer (thickness 2H = 120 $\mu$m), see Fig.~\ref{fig:Fig1}a. The bottom surface is patterned with cylindrical obstacles (radius 10 $\mu$m and height 6 $\mu$m), directly printed using two-photon polymerization lithography, see Appendix I. We connect the sample to a function generator through which we apply a transverse AC electric field of V$_{pp}$/2H, where V$_{pp}$ is the peak-to-peak voltage amplitude. We follow the colloids with an inverted microscope in real time at 10 frames per second under bright-field illumination. We then use custom tracking algorithms in Python to identify the locations of both particles and topographic features. 

At field frequencies between 100 Hz and 10 kHz, similar to the ones tested here, the colloids self-propel due to asymmetric electrodynamic flows generated along their surfaces~\cite{Squires2006, Ristenpart2007}. An ionic double layer is built near the conductive surfaces which act as electrodes. The electric field that establishes between them polarizes the particles and in turn induces an electric field gradient where the particle meets the bottom electrode, resulting in electrohydrodynamic flow fields~\cite{Ma2015, Yang2019}.
If the particles were compositionally symmetric, the flows would also be symmetric and therefore there would not be net motion. However, due to different polarizability of the metallic and dielectric sides of the particles, there exists an imbalance in the electrohydrodynamic flows between the two sides, causing self-propulsion with the metallic half at the back. Under the conditions applied in our experiments this leads to self-propulsion velocities ranging between 15 and 20 $\mu$m/s, see Fig. S1 in Appendix II.\\

We first examine the effect of a single cylindrical obstacle on approaching microswimmers, and find that they change their motion direction and turn away from the obstacle. This behavior stems from the fact that, in our experiments, flows are generated not only along the swimmer surfaces but also around the obstacles. The combined electrohydrodynamic flows effectively create a long-ranged repulsion between them. Example active trajectories that illustrate this effect are shown in Fig.~\ref{fig:Fig1}b-\ref{fig:Fig1}d. 

Specifically, at a frequency $f$ = 9.0~kHz, the particle ``senses'' the obstacle and avoids it (Fig.~\ref{fig:Fig1}b). We find that this persists for a range of frequencies in $f$ = 1.0~kHz (Fig.~\ref{fig:Fig1}c) and $f$ = 0.6~kHz (Fig.~\ref{fig:Fig1}d), albeit with ``sensing" and turning away taking place at increasingly larger distances. We observe that, for all frequencies, particles slow down as they swim towards the obstacle and subsequently reorient and swim away with a higher speed (see the trajectories in Fig.~\ref{fig:Fig1}, which we color-code according to the instantaneous velocity). Sufficiently far away from the obstacle, particles recover their frequency-dependent swim speed. 

\begin{figure*}[!hb]
    \centering
    \includegraphics[width=1\linewidth]{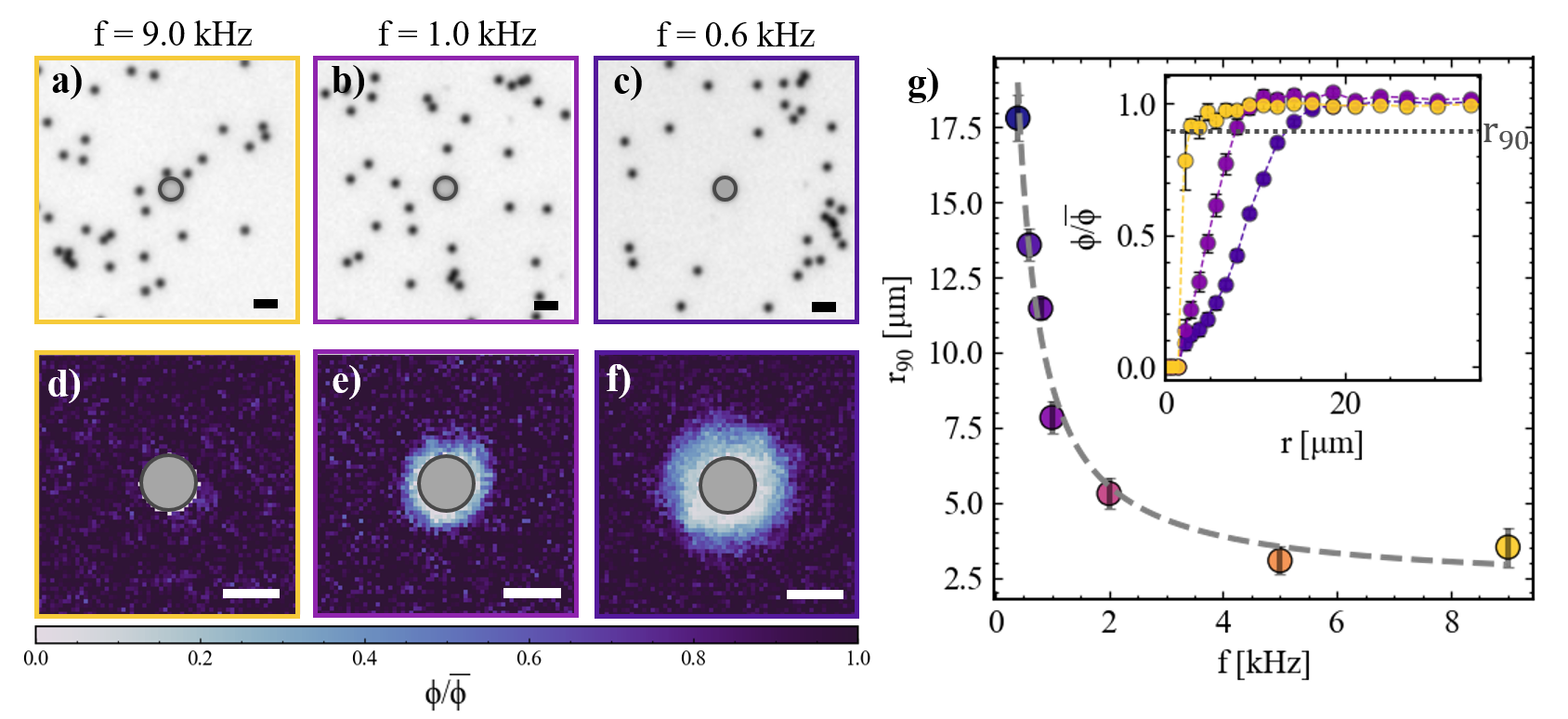}\vspace{-15pt}
    \caption{\textbf{Tunable long-ranged microswimmer-obstacle interaction through adjusting the applied field frequency, $f$.} Bright-field microscopy images with active colloids in proximity to a cylindrical obstacle (radius 10 $\mu$m, height 6 $\mu$m) under a fixed 6 V peak-to-peak amplitude and $f$ of a) 9.0~kHz, b) 1.0 kHz, and c) 0.6 kHz. d-e) Time-averaged particle density profiles around the obstacle corresponding to a-c, normalized by the time-averaged density in the sample far away from the obstacle. f) Particle-obstacle separation distance $r_{90}$, defined as the distance at which the normalized time-averaged microswimmer density is equal to 0.9, as a function of $f$. Inset shows the normalized particle density as a function of radial distance from the obstacle for $f$ = 0.6 kHz, 1.0 kHz, and 9.0 kHz. The dotted line indicates the extracted $r_{90}$ value. Scale bars are 10 $\mu$m.}
    \label{fig:Fig2}
\end{figure*}

Next, we seek to understand how the applied frequency affects particle dynamics. We quantify microswimmer-boundary interactions under various frequencies, see Fig.~\ref{fig:Fig2}a-\ref{fig:Fig2}c, by calculating time-averaged particle density profiles around the obstacles, see Appendix I and Fig.~\ref{fig:Fig2}d-\ref{fig:Fig2}f as examples; dark regions indicate areas of high probability of finding a particle and, \textit{vice versa}, light regions indicate areas of low probability.

At a frequency of $f$ = 9.0 kHz, we do not observe any apparent exclusion region around obstacles, as in Fig.~\ref{fig:Fig2}d. However, as we decrease the frequency of the electric field to 1.0 kHz in Fig.~\ref{fig:Fig2}e and 0.6 kHz in Fig.~\ref{fig:Fig2}f, a well-defined exclusion region appears, indicating an increase in the \textit{effective} size of the obstacle. Using the corresponding time-averaged density profiles, we calculate the normalized average particle density at each distance from the obstacle surface and at different frequencies. We show example curves that correspond to frequencies of $f$ = 9.0, 1.0 and 0.6 kHz in the inset of Fig.~\ref{fig:Fig2}g. We subsequently extract the distance from the obstacle where the time-averaged density profile equals 0.9, corresponding to a 90\% probability for a microswimmer to be able to reach the corresponding distance from the obstacle (represented by a horizontal dotted line, $r_{90}$, in the inset of Fig.~\ref{fig:Fig2}g). Plotting the extracted $r_{90}$ values as a function of $f$ in the main panel in turn reveals a monotonically-decreasing separation distance with increasing frequency. Detailed data at various frequencies within the $f$ = 0.6-9.0 kHz range are shown in Fig. S2 in Appendix II. 

The data in Fig.~\ref{fig:Fig2}g are well fitted with the expression $r_{90} = \frac{a}{f}+c$, with $a$ = 6.7 $\pm$ 0.15 and $c$ = 2.2 $\pm$ 0.17 $\mu$m for this specific sample (see dashed line). 
A similar behavior is found for bare 3 $\mu$m SiO$_2$ spheres, i.e. not self-propelling ones, albeit with a much larger exclusion zone compared to the Pd-capped ones, in Fig. S3 in Appendix II (note that self-organization of bare spheres in the passive state differs from that of the Janus spheres in the active state, especially in the low frequency regime where bare particles form large aggregates)~\cite{Ristenpart2007, Yang2019}. The inverse relationship between field frequency and the size of the exclusion region around the obstacle aligns well with theoretical predictions by Ristenpart \textit{et al.}~\cite{Ristenpart2004}, which suggest that the strength of electrohydrodynamic flows generated by a dielectric object (in our system the obstacle) near a conductive surface is inversely proportional to the frequency of the applied electric field. We hypothesize that active particles can approach the obstacles more closely than passive ones owing to their self-propulsion, which enables them to swim ``upstream" in the flows generated by the obstacles, while passive ones are simply pushed away and excluded by those flows.

\subsection*{\large Microswimmer gating in patterned environments}

The frequency-dependent effective obstacle size allows us to tune the behavior and self-organization of our active particles across different environments patterned with obstacles. We fabricate lattice configurations with different patterns, and vary both the spacing between neighboring obstacles as well as the effective obstacle size through frequency modulation, as established above. 

\begin{figure*}[t!]
    \centering
    \includegraphics[width=0.95\linewidth]{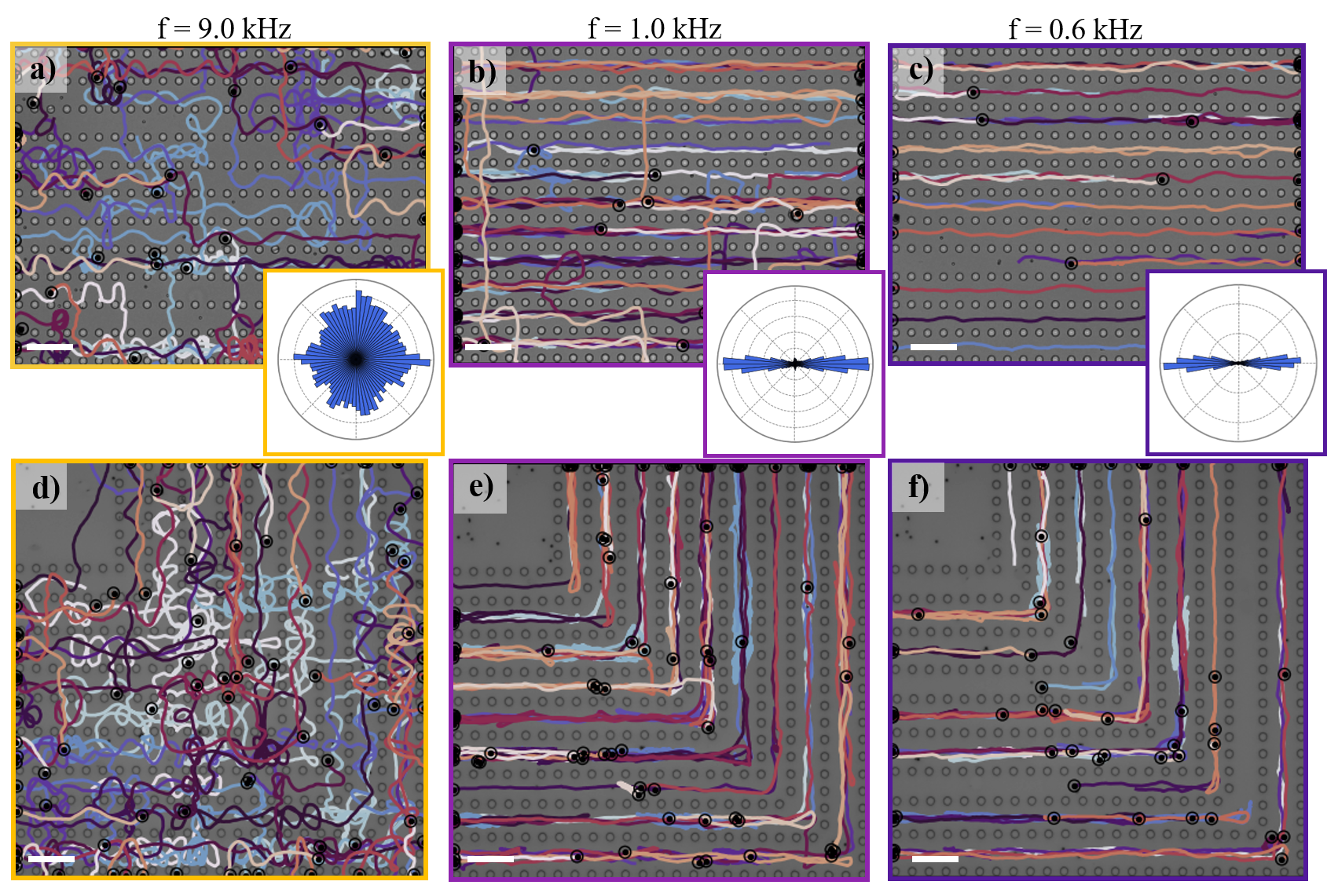}
    \vspace{-5pt}
   \caption{\textbf{Tunable microswimmer gating across environments with obstacles in straight and bent path configurations.} Representative active trajectories of colloids self-propelling through arrays of cylindrical obstacles with a lattice spacing of 10 $\mu$m in the x and 35 $\mu$m in the y direction, at varying applied frequency $f$ of a) 9.0 kHz, b) 1.0 kHz, and c) 0.6 kHz. Insets show the distributions of directions in the displacements of the particles over 1~s intervals, extracted from at least 60 trajectories of 2~min duration. Representative active trajectories of colloids self-propelling through an obstacle array with a lattice spacing of 10 $\mu$m in the x and 35 $\mu$m in the y direction constructed to form a path with a 90$^\circ$ bend, at $f$ of d)~9.0~kHz, e) 1.0 kHz, and f) 0.6 kHz. The peak-to-peak field amplitude is fixed in all experiments at 6 V. Scale bars are 50 $\mu$m. 
   }
    \label{fig:Fig3}
\end{figure*}

First, we examine active motion within straight arrays of obstacles with lattice spacing of 10 $\mu$m in the x direction and 35 $\mu$m in the y direction. We find that at a frequency $f$ = 9.0 kHz, particles move through the obstacle arrays without any preferred direction. In Fig.~\ref{fig:Fig3}a, we overlay example particle trajectories on top of a micrograph depicting the obstacle array and show that the active colloids are able to explore the environment unimpeded, as also evidenced by the distribution of directions in the particle displacements in the inset. As shown in Fig.~\ref{fig:Fig2}d, the effective obstacle size at this frequency is very similar to its physical dimensions and substantially smaller than the gaps between the obstacles, so the particles can easily swim between them. However, note that this case is different from the one of catalytically-active particles which are captured by obstacles~\cite{Baalen2023, Mijalkov2013, Katuri2018, Maggi2015,Reichhardt2017, Kaiser2012, Palacios2021, Takagi2014, Brown2016, Ketzetzi2022} .

At a frequency $f$ = 1 kHz, particles instead move predominantly in straight paths in the x direction parallel to the obstacle array, and rarely cross along in the y direction (Fig.~\ref{fig:Fig3}b). This behavior becomes even more pronounced at frequency $f$ = 0.6 kHz (Fig.~\ref{fig:Fig3}c), where active particles only appear to move along the x direction and in the middle of the obstacles, which effectively act as channels. This is further reflected in the distribution of directions, see the corresponding insets, as follows from Fig.~\ref{fig:Fig2}e-\ref{fig:Fig2}g. In those two cases, the separation between the obstacle columns is smaller than the effective obstacle size reported in Fig.~\ref{fig:Fig2}d, such that the gaps between obstacles in the y direction are effectively closed, while providing a guiding action in lanes in the x direction. Tuning the effective obstacle size thus allows us to gate active particle motion along specific directions. 

This frequency-dependent gating also enables guiding particles along more complex paths, i.e. around sharp bends. In Fig.~\ref{fig:Fig3}d-\ref{fig:Fig3}f, we add a 90$^\circ$ bend in the same lattice as previously described and, as before, we observe that at frequency $f$ = 9 kHz, our active colloids freely explore the obstacle lattice and actively distribute themselves across the entire available space (Fig.~\ref{fig:Fig3}d). Adjusting the frequency to $f$ = 1.0 kHz (Fig.~\ref{fig:Fig3}e) and to $f$ = 0.6 kHz (Fig.~\ref{fig:Fig3}f), again prevents particles from crossing across different rows in the lattice and forces them to follow obstacle lines, also around the bends, indicating that our frequency-dependent gating allows directing particles within complex environments. 

\newpage

\begin{figure*}[!hb]
\vspace{-30pt}
   \centering
    \includegraphics[width=1.0\linewidth]{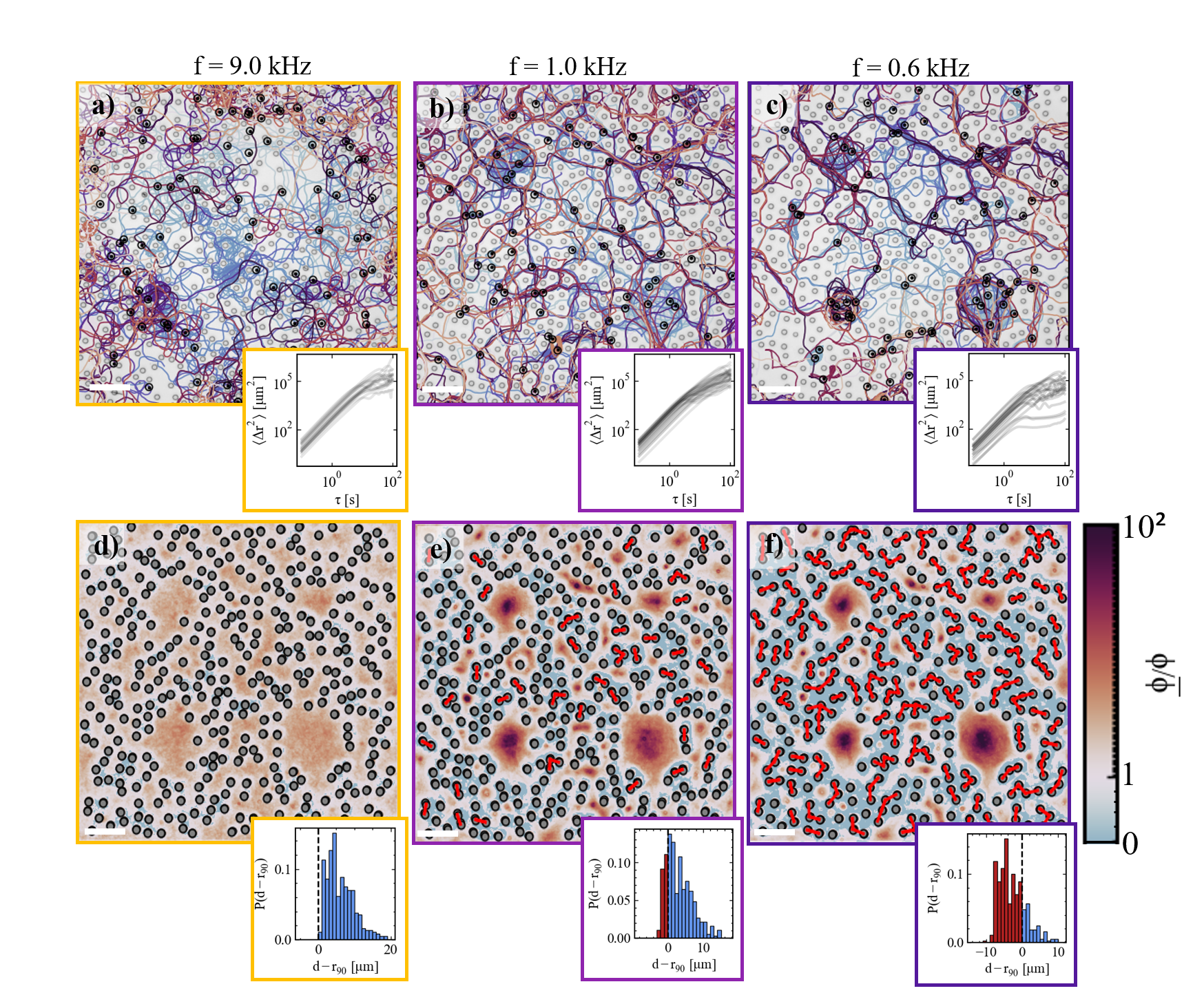}
    \vspace{-30pt}
    \caption{\textbf{Tunable confinement of synthetic microswimmers within disordered environments.} Representative trajectories of active colloids within a disordered array of 3D-printed obstacles under a fixed peak-to-peak voltage of $V_{pp}$ = 6 V and frequency $f$ of a) 9.0 kHz, b)~1.0 kHz, and c) 0.6 kHz. Trajectories are overlaid on top of a bright-field image of the patterned environment. Circles with black dots represent the active colloids in the last frame of the 2 min recording of their trajectories. Insets show corresponding mean squared displacements of individual colloids (IMSDs, $\langle \Delta r^2 \rangle$) as a function of lag time $\tau$. d-f) Time-averaged microswimmer density profiles corresponding to a-c). 
    Obstacles are plotted as light grey circles with dark grey edge lines. Changing the frequency increases the effective size of the obstacles creating closed paths, where the distance between obstacle surfaces $d$ is smaller than the corresponding $r_{90}$ value (marked by a red line connecting the obstacles). Insets show the distribution of nearest distance between obstacle surfaces, subtracted by the $r_{90}$ value of the corresponding frequency. Values of $d - r_{90}$ below 0 (black dashed line) correspond to closed paths, plotted as red bars in the distribution, whereas open paths are plotted as blue bars. Scale bars are 100 $\mu$m.}
    \label{fig:Fig4}
\end{figure*}

\clearpage

\subsection*{\large Tunable microswimmer confinement in disordered environments}

Finally, the gating strategy reported above gives us the opportunity to tune and directly control the confinement of active particles inside environments with a broad distribution of separations between obstacles, which we print here to form a 2D disordered environment. We show representative active trajectories obtained from 2 min recordings within these disordered environments in Fig.~\ref{fig:Fig4}a-\ref{fig:Fig4}c. Already from the trajectories, we see that tuning the applied field frequency leads to particle localization. 

At a frequency $f$ = 9.0 kHz, the trajectories of the swimmers approach the obstacles closely such that they can cross all gaps between the obstacles freely and explore all available space (Fig.~\ref{fig:Fig4}a). However, upon changing frequency to $f$ = 1.0 kHz in Fig.~\ref{fig:Fig4}b, a given fraction of gaps, i.e those for which the distance between obstacles becomes smaller than $r_{90}$, is effectively closed for the particles, which limits the regions they can explore. This effect becomes more pronounced when we lower the frequency further to $f$ = 0.6 kHz in Fig.~\ref{fig:Fig4}c. The occurrence of gating strongly influences active trajectories, as clearly shown by the mean square displacements of individual colloids as a function of lag time reported in the insets. In the presence of restrictions in space exploration, plateaus in the mean square displacements emerge as microswimmers become caged inside the larger cavities and cannot escape through the gaps between the obstacles. 

This frequency-dependent caging is clearly visualized by evaluating time-averaged microswimmer density maps under various applied frequencies, see Appendix I and Fig.~\ref{fig:Fig4}d-\ref{fig:Fig4}f, where each dataset represents an average over 16 measurements with a duration of 4 min. In particular, we mark closed gaps, which microswimmers cannot cross, with a red line that connects neighboring obstacles in Fig.~\ref{fig:Fig4}e and~\ref{fig:Fig4}f corresponding to $f$ = 1.0 and $f$ = 0.6 kHz, respectively. The number of closed paths in the system increases with decreasing frequency, as shown in the distribution reported in the inset. Specifically, the blue bars in the distribution correspond to open, permitted paths and red bars to closed ones in relation to the corresponding values of $r_{90}$.

\vspace{-5pt}
\subsection*{\Large Conclusions} \label{Disc}
In summary, we demonstrate that we can engineer long-range interactions between colloids self-propelling under an electric field and obstacles configured in prescribed paths forming complex tunable environments. As a result of these interactions, active colloids turn away from obstacles and avoid accumulation at boundaries, showing a distinctive difference from the classically-observed behavior for catalytic synthetic microswimmers. By varying the effective obstacle size we can achieve gating and confinement of active colloids via a modulation of the gaps between obstacles that can be easily opened and closed. 

The precise and rapid regulation strategy that we describe here offers high potential for the dynamic control of active particle gating, and bypasses key limitations of previous studies that utilized catalytically self-propelled colloids in topographically-patterned environments. In particular, by introducing feedback schemes that connect gating to real-time particle distributions, we envisage future work towards motion rectification, dynamic sorting of particles based on motility, and creation of tunable self-assembled active particle patterns. We therefore propose that electric fields open up exciting avenues for manipulating synthetic active matter in complex environments giving access to interaction modes with confinement geometries that expand existing possibilities towards the realization of advanced autonomous active units.

\subsection*{\Large Acknowledgements}
The authors thank Moran Bercovici and Robert Style for discussions. C.v.B. acknowledges funding from the European Union’s Horizon 2020 MSCA-ITN-ETN, project number 812780. L.I. and S.K. acknowledge funding from the European Research Council (ERC) under the European Union’s Horizon 2020 Research and innovation program grant agreement No 101001514.

\subsection*{\Large Author Contributions}
Author contributions are defined based on the CRediT (Contributor Roles Taxonomy) and listed alphabetically. Conceptualization: C.v.B, L.I, S.K. Formal Analysis: C.v.B., A.T. Funding acquisition: L.I. Investigation: C.v.B., S.K., A.T. Methodology: C.v.B., L.I., S.K. Project Administration: L.I. Software: C.v.B. Supervision: L.I., S.K. Validation: C.v.B., S.K. Visualization: C.v.B, L.I., S.K., A.T. Writing - original draft: C.v.B., S.K. Writing - review and editing: C.v.B, L.I., S.K.

\subsection*{\Large Appendix I: Materials and Methods}
\subsection{Janus particles}
Metallo-dielectric Janus colloids are fabricated by drop-casting an aqueous suspension of SiO$_2$ spheres (R=2.96 $\mu$m, SiO2-R-LSC84, microparticles GmbH) on a plasma-cleaned microscopy slide, followed by depositing a thin (2 nm) Cr adhesion layer and a 6.5 nm Pd layer. The resulting Janus spheres were dispersed in 50 mL Milli-Q water, and washed 3 times by centrifugation and redispersion in fresh Milli-Q water. Finally, the particles were concentrated to a volume of 0.5 mL to obtain a suspension of approximately 2 mg particles per mL solution.

\subsection{Patterned environments}
Microstructures were produced with a Nanoscribe Photonic Professional GT2, which uses two-photon lithography. Designs comprised cylindrical obstacles with height 6 $\mu$m and diameter 10 $\mu$m in various configurations, and were designed using a CAD software (Autodesk Fusion) and processed with Describe. Obstacles were printed onto UV-ozone treated ITO-coated glass slides (Nanoscribe) using the commercial photoresist IP-S as a pre-polymer. Standard printing parameters were selected, as specified by the manufacturer: the printer was equipped with a 25x-immersion objective (Zeiss, NA = 0.8) and used to print in DiLL mode. After printing, the structures were developed by submersion in propylene glycol methylether acrylate for 15 min, immediately followed by gently dipping into isopropyl alcohol and gentle drying with a nitrogen gun. This procedure reliable removes the unpolymerized photoresist. All micro-patterning steps were performed under yellow light. Finally, the printed structures were post-cured for 6 min under a UV lamp with wavelength 565 nm.

\subsection{Preparation of the experimental cell}
Prior to the experiments a small amount of the Janus particle suspension was mixed in a 1:1 ratio with a 2\% surfactant (Pluronic-127, Sigma Aldrich) solution. A 5 $\mu$L droplet of the suspension was placed in a custom-made sample cell consisting of two transparent electrodes separated by an adhesive spacer with a 9 mm-circular opening and 120 $\mu$m height (Grace Bio-Labs SecureSeal). The bottom electrodes were decorated with obstacles as described above and the top ones were plain ITO-coated slides cleaned via 20 min sonication in acetone and Milli-Q water, followed by drying with a pure nitrogen stream before assembling the sample cell. Once the particles were added and the electrodes adhered via the spacer, the electrodes were connected using copper tape to a function generator (National Instruments Agilent 3352X) that applies the AC electric field ($f$= 0.6 - 9.0 kHz, $V_{pp}$ = 6 V). 

\subsection{Imaging}
The sample cell was mounted on an Eclipse Ti2 inverted microscope in bright-field mode and videos were recorded at a frame rate of 10 frames per second using a CMOS camera (Orca Flash 4.0 V3, Hammamatsu). Data for partice density mapping were obtained using a 4x objective (Plan Fluor 4x, Nikon, NA = 0.13) with 1.5 magnification lens. In addition, higher magnification data were obtained for particle tracking and Janus cap visualization using 20x (S Plan Fluor 20x, Nikon, NA = 0.45) and 40x (S Plan Fluor 40x, Nikon, NA = 0.60) objectives. 

\subsection{Particle tracking}
The acquired images were preprocessed in Python by inverting their intensity values and binarizing them to enhance the contrast between obstacles and the background. The centers of the obstacles were identified in the binarized images using OpenCV. Obstacle detection was achieved by applying an area threshold to select features within the desired size range. Once identified, obstacle regions were masked out from the images to exclude them from further analysis. Particle tracking was then performed on the remaining masked images using a centroid finding algorithm implemented in Python (Trackpy)~\cite{Crocker1996, Trackpy}.

\subsection{Time-averaged density maps around single obstacles}
For the time-averaged density maps around obstacles, we followed the same preprocessing procedure as described above. However, instead of tracking the particles, the masked images were overlaid. To obtain the density profiles shown in Fig. 2 as well as Fig. S2, and Fig. S3, 16 image sequences with a field of view of 102 × 102 pixels, each lasting 2 minutes and recorded at 10 fps, were overlaid. The images were normalized by the mean pixel value calculated from the 10 pixels along the edges of the time-averaged image. For the density profiles shown in Fig. 4, data from 16 experiments, each lasting 4 minutes with a field of view of 550 × 550 pixels, were overlaid based on the best alignment of the obstacle centers in the corresponding image sequences. The resulting density maps were normalized by the time-average pixel intensity measured 50-55 $\mu$m away from the disordered lattice.

\setcounter{figure}{0}
\newcommand{\bluec}[1]{\textcolor{blue}{[#1]}}
\newcommand{\redc}[1]{\textcolor{red}{[#1]}}
\renewcommand{\thefigure}{S\arabic{figure}}

\subsection*{\Large Appendix II: Supporting data}

\begin{figure}[h!]
    \centering
    \includegraphics[width=14cm]{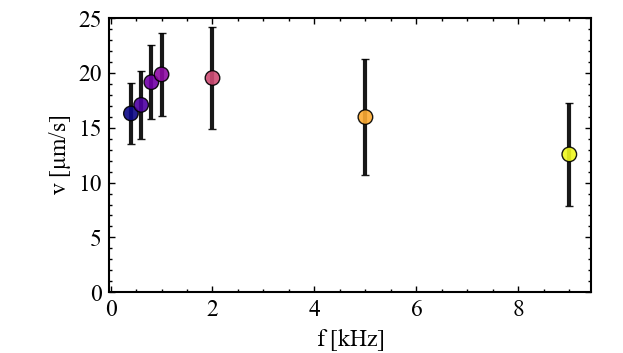}
    \caption{\textbf{Microswimmer self-propulsion speed under an electric field.} Active velocities of metallodielectric Janus spheres composed of a 3 $\mu$m SiO$_2$ core and a 6.5 nm Pd cap as a function of the frequency of the applied alternating current (AC) electric field. Data was obtained at a fixed peak-to-peak voltage of $V_{pp}$ = 6 V. Self-propulsion speeds were extracted from a least-squares fit of the individual mean squared displacements of approximately 60 microswimmers. }
    \label{fig:SI1}
\end{figure}

\begin{figure}[h!]
    \centering
    \includegraphics[width=16cm]{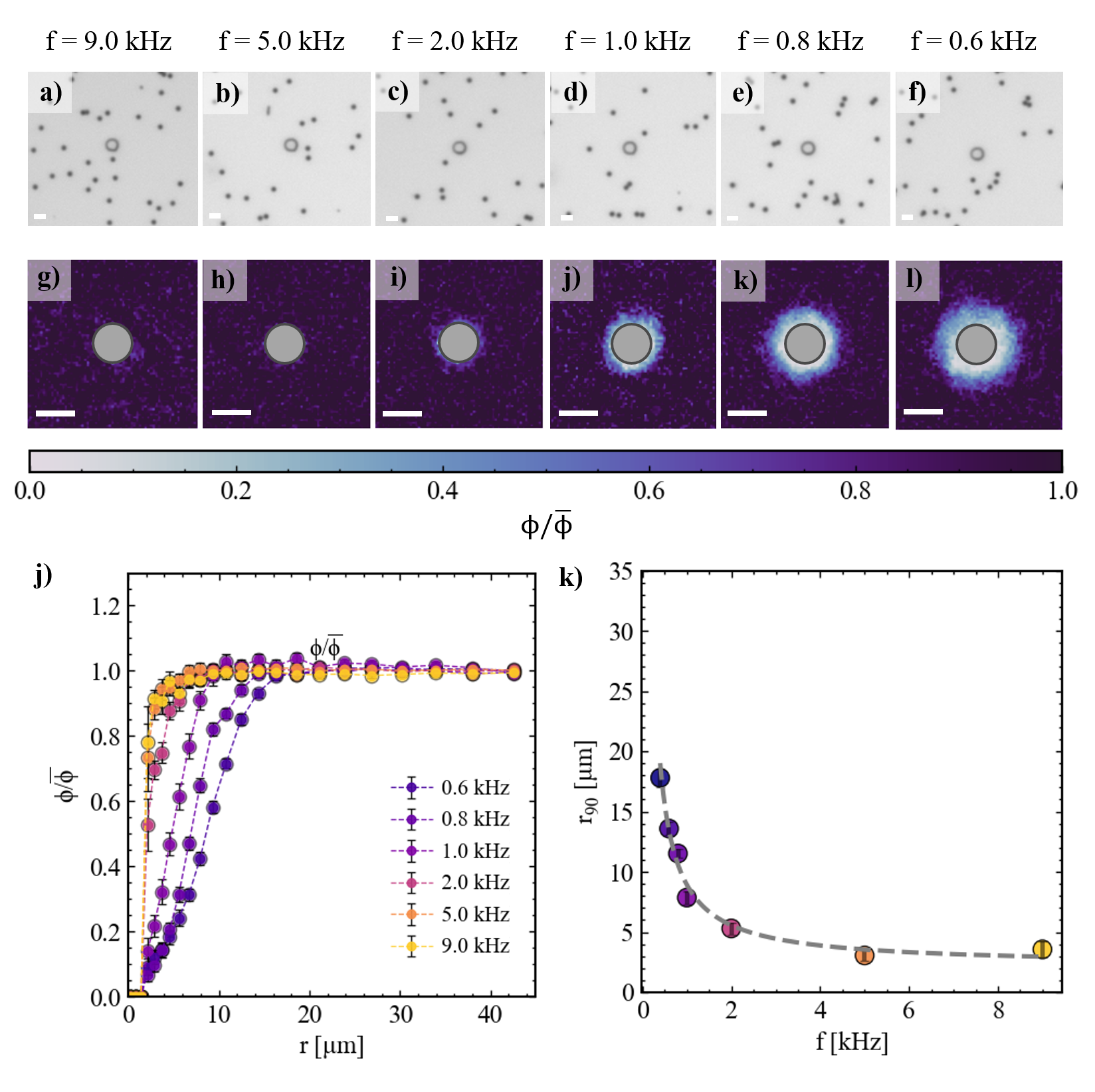}
    \vspace{-20pt}
    \caption{\textbf{Detailed data of microswimmer obstacle avoidance via electric field frequency modulation (active state).} a-f) Bright-field microscopy images of a sample depicting Janus Pd-capped active colloids in proximity to a cylindrical obstacle with height 6 $\mu$m and radius 10 $\mu$m under a fixed 6 V peak-to-peak amplitude and varying frequency from $f$ = 9.0 - 0.6 kHz. g-l) Time-averaged Janus microswimmer density profiles corresponding to a-f, normalized by the time-averaged density in the sample far away from the obstacle. j) Normalized microswimmer density as a function of radial distance from the obstacle surface for frequencies ranging from $f$ = 9.0 - 0.6 kHz. k) Microswimmer-obstacle separation distance at a time-averaged microswimmer density corresponding to 0.9 times the density far away from the obstacle, i.e. $r_{90}$, as a function of the frequency. Scale bars are 10 $\mu$m.}
    \label{fig:SI2}
\end{figure}

\begin{figure}[h!]
    \centering
    \includegraphics[width=16cm]{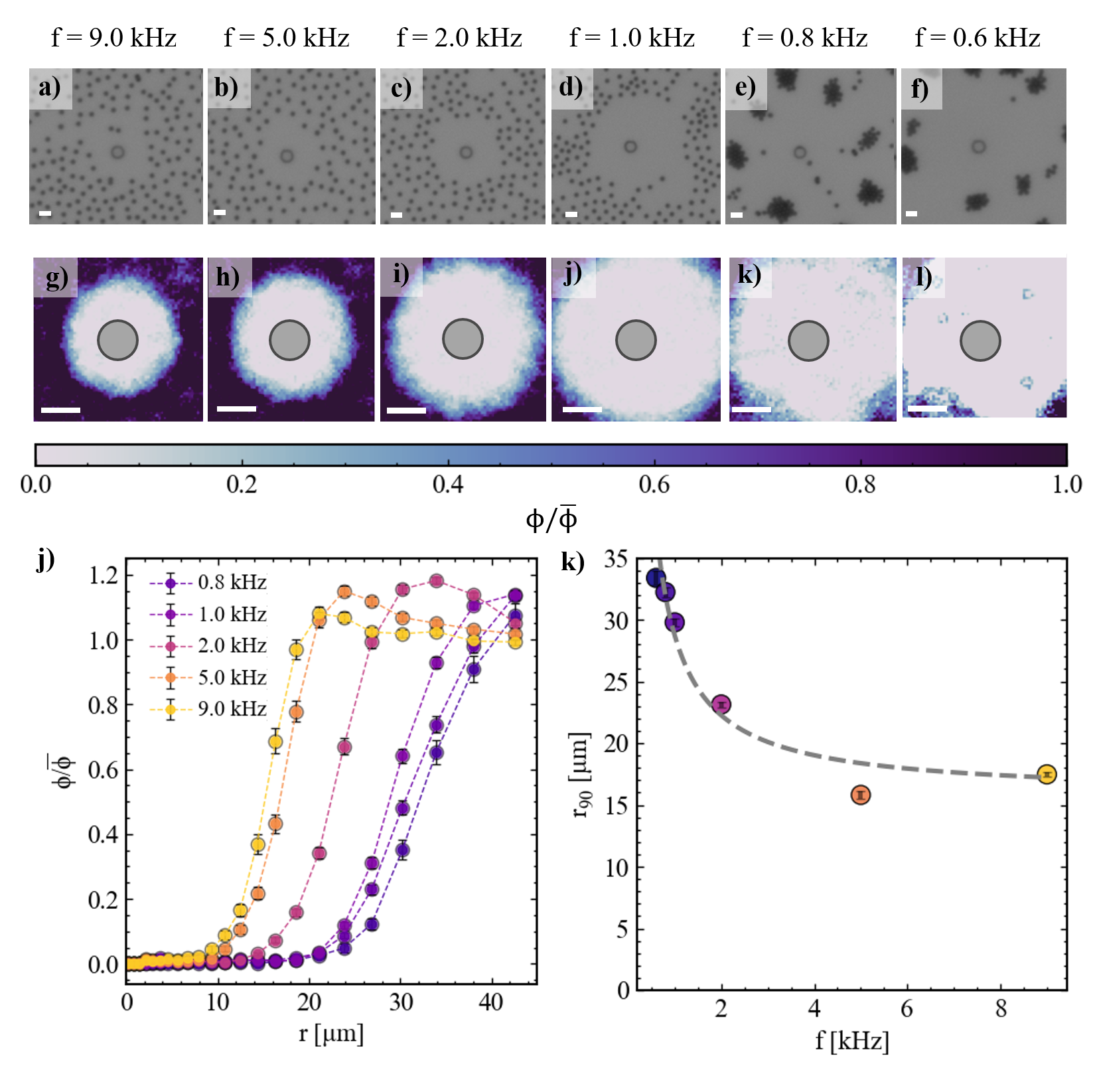}
    \vspace{-28pt}
    \caption{\textbf{Detailed data of bare SiO$_2$ colloids as a function of the electric field frequency (passive state).} a-f) Bright-field microscopy images of a sample depicting how bare SiO$_2$ colloids distribute and self-organize in proximity to a cylindrical obstacle with height 6 $\mu$m and radius 10 $\mu$m under a fixed 6 V peak-to-peak amplitude and varying frequency. At frequencies between 9.0-1.0 kHz, the colloids remain far from the obstacle and each other, while at frequencies below 1.0 kHz they form aggregates, differently from the Pd-capped particles in the active state in Fig. S2. g-l) Time-averaged particle density profiles corresponding to a-f, normalized by the time-averaged density in the sample far away from the obstacle. j) Normalized particle density as a function of radial distance from the obstacle surface for the same frequencies. k) Particle-obstacle separation distance at a time-averaged particle density corresponding to 0.9 times the density far from the obstacle, i.e. $r_{90}$, with frequency. Scale bars are 10 $\mu$m.}
    \label{fig:SI3}
\end{figure}

\clearpage

\bibliography{Main}

\begin{thebibliography}{62}%
\makeatletter
\providecommand \@ifxundefined [1]{%
 \@ifx{#1\undefined}
}%
\providecommand \@ifnum [1]{%
 \ifnum #1\expandafter \@firstoftwo
 \else \expandafter \@secondoftwo
 \fi
}%
\providecommand \@ifx [1]{%
 \ifx #1\expandafter \@firstoftwo
 \else \expandafter \@secondoftwo
 \fi
}%
\providecommand \natexlab [1]{#1}%
\providecommand \enquote  [1]{``#1''}%
\providecommand \bibnamefont  [1]{#1}%
\providecommand \bibfnamefont [1]{#1}%
\providecommand \citenamefont [1]{#1}%
\providecommand \href@noop [0]{\@secondoftwo}%
\providecommand \href [0]{\begingroup \@sanitize@url \@href}%
\providecommand \@href[1]{\@@startlink{#1}\@@href}%
\providecommand \@@href[1]{\endgroup#1\@@endlink}%
\providecommand \@sanitize@url [0]{\catcode `\\12\catcode `\$12\catcode `\&12\catcode `\#12\catcode `\^12\catcode `\_12\catcode `\%12\relax}%
\providecommand \@@startlink[1]{}%
\providecommand \@@endlink[0]{}%
\providecommand \url  [0]{\begingroup\@sanitize@url \@url }%
\providecommand \@url [1]{\endgroup\@href {#1}{\urlprefix }}%
\providecommand \urlprefix  [0]{URL }%
\providecommand \Eprint [0]{\href }%
\providecommand \doibase [0]{http://dx.doi.org/}%
\providecommand \selectlanguage [0]{\@gobble}%
\providecommand \bibinfo  [0]{\@secondoftwo}%
\providecommand \bibfield  [0]{\@secondoftwo}%
\providecommand \translation [1]{[#1]}%
\providecommand \BibitemOpen [0]{}%
\providecommand \bibitemStop [0]{}%
\providecommand \bibitemNoStop [0]{.\EOS\space}%
\providecommand \EOS [0]{\spacefactor3000\relax}%
\providecommand \BibitemShut  [1]{\csname bibitem#1\endcsname}%
\let\auto@bib@innerbib\@empty
\bibitem [{\citenamefont {Moreno-Gámez}\ \emph {et~al.}(2017)\citenamefont {Moreno-Gámez}, \citenamefont {Sorg}, \citenamefont {Domenech}, \citenamefont {Kjos}, \citenamefont {Weissing}, \citenamefont {van Doorn},\ and\ \citenamefont {Veening}}]{Moreno2017}%
  \BibitemOpen
  \bibfield  {author} {\bibinfo {author} {\bibfnamefont {S.}~\bibnamefont {Moreno-Gámez}}, \bibinfo {author} {\bibfnamefont {R.~A.}\ \bibnamefont {Sorg}}, \bibinfo {author} {\bibfnamefont {A.}~\bibnamefont {Domenech}}, \bibinfo {author} {\bibfnamefont {M.}~\bibnamefont {Kjos}}, \bibinfo {author} {\bibfnamefont {F.~J.}\ \bibnamefont {Weissing}}, \bibinfo {author} {\bibfnamefont {G.~S.}\ \bibnamefont {van Doorn}}, \ and\ \bibinfo {author} {\bibfnamefont {J.-W.}\ \bibnamefont {Veening}},\ }\href@noop {} {\bibfield  {journal} {\bibinfo  {journal} {Nat Commun}\ }\textbf {\bibinfo {volume} {8}},\ \bibinfo {pages} {854} (\bibinfo {year} {2017})}\BibitemShut {NoStop}%
\bibitem [{\citenamefont {Lele}\ \emph {et~al.}(2013)\citenamefont {Lele}, \citenamefont {Hosu},\ and\ \citenamefont {Berg}}]{Lele2013}%
  \BibitemOpen
  \bibfield  {author} {\bibinfo {author} {\bibfnamefont {P.~P.}\ \bibnamefont {Lele}}, \bibinfo {author} {\bibfnamefont {B.~G.}\ \bibnamefont {Hosu}}, \ and\ \bibinfo {author} {\bibfnamefont {H.~C.}\ \bibnamefont {Berg}},\ }\href@noop {} {\bibfield  {journal} {\bibinfo  {journal} {Proc. Natl. Acad. Sci.}\ }\textbf {\bibinfo {volume} {110}},\ \bibinfo {pages} {11839} (\bibinfo {year} {2013})}\BibitemShut {NoStop}%
\bibitem [{\citenamefont {Persat}\ \emph {et~al.}(2015)\citenamefont {Persat}, \citenamefont {Nadell}, \citenamefont {Kim}, \citenamefont {Ingremeau}, \citenamefont {Siryaporn}, \citenamefont {Drescher}, \citenamefont {Wingreen}, \citenamefont {Bassler}, \citenamefont {Gitai},\ and\ \citenamefont {Stone}}]{Persat2015}%
  \BibitemOpen
  \bibfield  {author} {\bibinfo {author} {\bibfnamefont {A.}~\bibnamefont {Persat}}, \bibinfo {author} {\bibfnamefont {C.~D.}\ \bibnamefont {Nadell}}, \bibinfo {author} {\bibfnamefont {M.~K.}\ \bibnamefont {Kim}}, \bibinfo {author} {\bibfnamefont {F.}~\bibnamefont {Ingremeau}}, \bibinfo {author} {\bibfnamefont {A.}~\bibnamefont {Siryaporn}}, \bibinfo {author} {\bibfnamefont {K.}~\bibnamefont {Drescher}}, \bibinfo {author} {\bibfnamefont {N.~S.}\ \bibnamefont {Wingreen}}, \bibinfo {author} {\bibfnamefont {B.~L.}\ \bibnamefont {Bassler}}, \bibinfo {author} {\bibfnamefont {Z.}~\bibnamefont {Gitai}}, \ and\ \bibinfo {author} {\bibfnamefont {H.~A.}\ \bibnamefont {Stone}},\ }\href@noop {} {\bibfield  {journal} {\bibinfo  {journal} {Cell}\ }\textbf {\bibinfo {volume} {161}},\ \bibinfo {pages} {988} (\bibinfo {year} {2015})}\BibitemShut {NoStop}%
\bibitem [{\citenamefont {Laventie}\ and\ \citenamefont {Jenal}(2020)}]{Laventie2020}%
  \BibitemOpen
  \bibfield  {author} {\bibinfo {author} {\bibfnamefont {B.-J.}\ \bibnamefont {Laventie}}\ and\ \bibinfo {author} {\bibfnamefont {U.}~\bibnamefont {Jenal}},\ }\href@noop {} {\bibfield  {journal} {\bibinfo  {journal} {Annu. Rev. Microbiol.}\ }\textbf {\bibinfo {volume} {74}},\ \bibinfo {pages} {735} (\bibinfo {year} {2020})}\BibitemShut {NoStop}%
\bibitem [{\citenamefont {Singh}\ \emph {et~al.}(2021)\citenamefont {Singh}, \citenamefont {Pagulayan}, \citenamefont {Camley},\ and\ \citenamefont {Nain}}]{Singh2021}%
  \BibitemOpen
  \bibfield  {author} {\bibinfo {author} {\bibfnamefont {J.}~\bibnamefont {Singh}}, \bibinfo {author} {\bibfnamefont {A.}~\bibnamefont {Pagulayan}}, \bibinfo {author} {\bibfnamefont {B.~A.}\ \bibnamefont {Camley}}, \ and\ \bibinfo {author} {\bibfnamefont {A.~S.}\ \bibnamefont {Nain}},\ }\href@noop {} {\bibfield  {journal} {\bibinfo  {journal} {PNAS}\ }\textbf {\bibinfo {volume} {118}},\ \bibinfo {pages} {e2011815118} (\bibinfo {year} {2021})}\BibitemShut {NoStop}%
\bibitem [{\citenamefont {Lin}\ \emph {et~al.}(2015)\citenamefont {Lin}, \citenamefont {Yin}, \citenamefont {Wu}, \citenamefont {Inoue},\ and\ \citenamefont {Levchenko}}]{Lin2015}%
  \BibitemOpen
  \bibfield  {author} {\bibinfo {author} {\bibfnamefont {B.}~\bibnamefont {Lin}}, \bibinfo {author} {\bibfnamefont {T.}~\bibnamefont {Yin}}, \bibinfo {author} {\bibfnamefont {Y.~I.}\ \bibnamefont {Wu}}, \bibinfo {author} {\bibfnamefont {T.}~\bibnamefont {Inoue}}, \ and\ \bibinfo {author} {\bibfnamefont {A.}~\bibnamefont {Levchenko}},\ }\href@noop {} {\bibfield  {journal} {\bibinfo  {journal} {Nat Commun}\ }\textbf {\bibinfo {volume} {6}},\ \bibinfo {pages} {6619} (\bibinfo {year} {2015})}\BibitemShut {NoStop}%
\bibitem [{\citenamefont {Hall-Stoodley}\ \emph {et~al.}(2004)\citenamefont {Hall-Stoodley}, \citenamefont {Costerton},\ and\ \citenamefont {Stoodley}}]{Stoodley2004}%
  \BibitemOpen
  \bibfield  {author} {\bibinfo {author} {\bibfnamefont {L.}~\bibnamefont {Hall-Stoodley}}, \bibinfo {author} {\bibfnamefont {J.~W.}\ \bibnamefont {Costerton}}, \ and\ \bibinfo {author} {\bibfnamefont {P.}~\bibnamefont {Stoodley}},\ }\href {\doibase 10.1038/nrmicro821} {\bibfield  {journal} {\bibinfo  {journal} {Nat Rev Microbiol}\ }\textbf {\bibinfo {volume} {2}},\ \bibinfo {pages} {95} (\bibinfo {year} {2004})}\BibitemShut {NoStop}%
\bibitem [{\citenamefont {Lee}\ \emph {et~al.}(2020)\citenamefont {Lee}, \citenamefont {Vachier}, \citenamefont {de~Anda}, \citenamefont {Zhao}, \citenamefont {Baker}, \citenamefont {Bennett}, \citenamefont {Armbruster}, \citenamefont {Lewis}, \citenamefont {Tarnopol}, \citenamefont {Lomba}, \citenamefont {Hogan}, \citenamefont {Parsek}, \citenamefont {O'Toole}, \citenamefont {Golestanian},\ and\ \citenamefont {Wong}}]{Lee2020}%
  \BibitemOpen
  \bibfield  {author} {\bibinfo {author} {\bibfnamefont {C.~K.}\ \bibnamefont {Lee}}, \bibinfo {author} {\bibfnamefont {J.}~\bibnamefont {Vachier}}, \bibinfo {author} {\bibfnamefont {J.}~\bibnamefont {de~Anda}}, \bibinfo {author} {\bibfnamefont {K.}~\bibnamefont {Zhao}}, \bibinfo {author} {\bibfnamefont {A.~E.}\ \bibnamefont {Baker}}, \bibinfo {author} {\bibfnamefont {R.~R.}\ \bibnamefont {Bennett}}, \bibinfo {author} {\bibfnamefont {C.~R.}\ \bibnamefont {Armbruster}}, \bibinfo {author} {\bibfnamefont {K.~A.}\ \bibnamefont {Lewis}}, \bibinfo {author} {\bibfnamefont {R.~L.}\ \bibnamefont {Tarnopol}}, \bibinfo {author} {\bibfnamefont {C.~J.}\ \bibnamefont {Lomba}}, \bibinfo {author} {\bibfnamefont {D.~A.}\ \bibnamefont {Hogan}}, \bibinfo {author} {\bibfnamefont {M.~R.}\ \bibnamefont {Parsek}}, \bibinfo {author} {\bibfnamefont {G.~A.}\ \bibnamefont {O'Toole}}, \bibinfo {author} {\bibfnamefont {R.}~\bibnamefont {Golestanian}}, \ and\ \bibinfo {author} {\bibfnamefont {G.~C.~L.}\ \bibnamefont {Wong}},\ }\href@noop
  {} {\bibfield  {journal} {\bibinfo  {journal} {mBio}\ }\textbf {\bibinfo {volume} {11}},\ \bibinfo {pages} {1} (\bibinfo {year} {2020})}\BibitemShut {NoStop}%
\bibitem [{\citenamefont {DiLuzio}\ \emph {et~al.}(2005)\citenamefont {DiLuzio}, \citenamefont {Turner}, \citenamefont {Mayer}, \citenamefont {Garstecki}, \citenamefont {Weibel}, \citenamefont {Berg},\ and\ \citenamefont {Whitesides}}]{DiLuzio2005}%
  \BibitemOpen
  \bibfield  {author} {\bibinfo {author} {\bibfnamefont {W.~R.}\ \bibnamefont {DiLuzio}}, \bibinfo {author} {\bibfnamefont {L.}~\bibnamefont {Turner}}, \bibinfo {author} {\bibfnamefont {M.}~\bibnamefont {Mayer}}, \bibinfo {author} {\bibfnamefont {P.}~\bibnamefont {Garstecki}}, \bibinfo {author} {\bibfnamefont {D.~B.}\ \bibnamefont {Weibel}}, \bibinfo {author} {\bibfnamefont {H.~C.}\ \bibnamefont {Berg}}, \ and\ \bibinfo {author} {\bibfnamefont {G.~M.}\ \bibnamefont {Whitesides}},\ }\href {\doibase 10.1038/nature03660} {\bibfield  {journal} {\bibinfo  {journal} {Nature}\ }\textbf {\bibinfo {volume} {435}},\ \bibinfo {pages} {1271} (\bibinfo {year} {2005})}\BibitemShut {NoStop}%
\bibitem [{\citenamefont {Drescher}\ \emph {et~al.}(2011)\citenamefont {Drescher}, \citenamefont {Dunkel}, \citenamefont {Cisneros}, \citenamefont {Ganguly},\ and\ \citenamefont {Goldstein}}]{Drescher2011}%
  \BibitemOpen
  \bibfield  {author} {\bibinfo {author} {\bibfnamefont {K.}~\bibnamefont {Drescher}}, \bibinfo {author} {\bibfnamefont {J.}~\bibnamefont {Dunkel}}, \bibinfo {author} {\bibfnamefont {L.~H.}\ \bibnamefont {Cisneros}}, \bibinfo {author} {\bibfnamefont {S.}~\bibnamefont {Ganguly}}, \ and\ \bibinfo {author} {\bibfnamefont {R.~E.}\ \bibnamefont {Goldstein}},\ }\href {\doibase 10.1073/pnas.1019079108} {\bibfield  {journal} {\bibinfo  {journal} {PNAS}\ }\textbf {\bibinfo {volume} {108}},\ \bibinfo {pages} {10940} (\bibinfo {year} {2011})}\BibitemShut {NoStop}%
\bibitem [{\citenamefont {Kantsler}\ \emph {et~al.}(2013)\citenamefont {Kantsler}, \citenamefont {Dunkel}, \citenamefont {Polin},\ and\ \citenamefont {Goldstein}}]{Kantsler2013}%
  \BibitemOpen
  \bibfield  {author} {\bibinfo {author} {\bibfnamefont {V.}~\bibnamefont {Kantsler}}, \bibinfo {author} {\bibfnamefont {J.}~\bibnamefont {Dunkel}}, \bibinfo {author} {\bibfnamefont {M.}~\bibnamefont {Polin}}, \ and\ \bibinfo {author} {\bibfnamefont {R.~E.}\ \bibnamefont {Goldstein}},\ }\href {\doibase 10.1073/pnas.1210548110} {\bibfield  {journal} {\bibinfo  {journal} {PNAS}\ }\textbf {\bibinfo {volume} {110}},\ \bibinfo {pages} {1187} (\bibinfo {year} {2013})}\BibitemShut {NoStop}%
\bibitem [{\citenamefont {Berne}\ \emph {et~al.}(2016)\citenamefont {Berne}, \citenamefont {Ellison},\ and\ \citenamefont {Ducret}}]{Berne2016}%
  \BibitemOpen
  \bibfield  {author} {\bibinfo {author} {\bibfnamefont {C.}~\bibnamefont {Berne}}, \bibinfo {author} {\bibfnamefont {C.~K.}\ \bibnamefont {Ellison}}, \ and\ \bibinfo {author} {\bibfnamefont {A.}~\bibnamefont {Ducret}},\ }\href {\doibase 10.1038/s41579-018-0057-5} {\bibfield  {journal} {\bibinfo  {journal} {Nat Rev Microbiol}\ }\textbf {\bibinfo {volume} {18}},\ \bibinfo {pages} {616} (\bibinfo {year} {2016})}\BibitemShut {NoStop}%
\bibitem [{\citenamefont {Lauga}\ and\ \citenamefont {Powers}(2009)}]{Lauga2009}%
  \BibitemOpen
  \bibfield  {author} {\bibinfo {author} {\bibfnamefont {E.}~\bibnamefont {Lauga}}\ and\ \bibinfo {author} {\bibfnamefont {T.~R.}\ \bibnamefont {Powers}},\ }\href@noop {} {\bibfield  {journal} {\bibinfo  {journal} {Rep. Prog. Phys.}\ }\textbf {\bibinfo {volume} {72}},\ \bibinfo {pages} {096601} (\bibinfo {year} {2009})}\BibitemShut {NoStop}%
\bibitem [{\citenamefont {Elgeti}\ \emph {et~al.}(2015)\citenamefont {Elgeti}, \citenamefont {Winkler},\ and\ \citenamefont {Gompper}}]{Elgeti2015}%
  \BibitemOpen
  \bibfield  {author} {\bibinfo {author} {\bibfnamefont {J.}~\bibnamefont {Elgeti}}, \bibinfo {author} {\bibfnamefont {R.~G.}\ \bibnamefont {Winkler}}, \ and\ \bibinfo {author} {\bibfnamefont {G.}~\bibnamefont {Gompper}},\ }\href {\doibase 10.1088/0034-4885/78/5/056601} {\bibfield  {journal} {\bibinfo  {journal} {Rep Prog Phys}\ }\textbf {\bibinfo {volume} {78}},\ \bibinfo {pages} {056601} (\bibinfo {year} {2015})}\BibitemShut {NoStop}%
\bibitem [{\citenamefont {Dreyfus}\ \emph {et~al.}(2005)\citenamefont {Dreyfus}, \citenamefont {Baudry}, \citenamefont {Roper}, \citenamefont {Fermigier}, \citenamefont {Stone},\ and\ \citenamefont {Bibette}}]{Dreyfus2005}%
  \BibitemOpen
  \bibfield  {author} {\bibinfo {author} {\bibfnamefont {R.}~\bibnamefont {Dreyfus}}, \bibinfo {author} {\bibfnamefont {J.}~\bibnamefont {Baudry}}, \bibinfo {author} {\bibfnamefont {M.~L.}\ \bibnamefont {Roper}}, \bibinfo {author} {\bibfnamefont {M.}~\bibnamefont {Fermigier}}, \bibinfo {author} {\bibfnamefont {H.~A.}\ \bibnamefont {Stone}}, \ and\ \bibinfo {author} {\bibfnamefont {J.}~\bibnamefont {Bibette}},\ }\href@noop {} {\bibfield  {journal} {\bibinfo  {journal} {Nature}\ }\textbf {\bibinfo {volume} {437}},\ \bibinfo {pages} {862} (\bibinfo {year} {2005})}\BibitemShut {NoStop}%
\bibitem [{\citenamefont {Golestanian}\ \emph {et~al.}(2005)\citenamefont {Golestanian}, \citenamefont {Liverpool},\ and\ \citenamefont {Ajdari}}]{Golestanian2005}%
  \BibitemOpen
  \bibfield  {author} {\bibinfo {author} {\bibfnamefont {R.}~\bibnamefont {Golestanian}}, \bibinfo {author} {\bibfnamefont {T.~B.}\ \bibnamefont {Liverpool}}, \ and\ \bibinfo {author} {\bibfnamefont {A.}~\bibnamefont {Ajdari}},\ }\href {\doibase 10.1103/PhysRevLett.94.220801} {\bibfield  {journal} {\bibinfo  {journal} {Phys. Rev. Lett.}\ }\textbf {\bibinfo {volume} {94}},\ \bibinfo {pages} {220801} (\bibinfo {year} {2005})}\BibitemShut {NoStop}%
\bibitem [{\citenamefont {Bechinger}\ \emph {et~al.}(2016)\citenamefont {Bechinger}, \citenamefont {Leonardo}, \citenamefont {L{\"o}wen}, \citenamefont {Reichhardt}, \citenamefont {Volpe},\ and\ \citenamefont {Volpe}}]{Bechinger2016}%
  \BibitemOpen
  \bibfield  {author} {\bibinfo {author} {\bibfnamefont {C.}~\bibnamefont {Bechinger}}, \bibinfo {author} {\bibfnamefont {R.~D.}\ \bibnamefont {Leonardo}}, \bibinfo {author} {\bibfnamefont {H.}~\bibnamefont {L{\"o}wen}}, \bibinfo {author} {\bibfnamefont {C.}~\bibnamefont {Reichhardt}}, \bibinfo {author} {\bibfnamefont {G.}~\bibnamefont {Volpe}}, \ and\ \bibinfo {author} {\bibfnamefont {G.}~\bibnamefont {Volpe}},\ }\href {\doibase 10.1103/RevModPhys.88.045006} {\bibfield  {journal} {\bibinfo  {journal} {Rev. Mod. Phys.}\ }\textbf {\bibinfo {volume} {88}},\ \bibinfo {pages} {045006} (\bibinfo {year} {2016})}\BibitemShut {NoStop}%
\bibitem [{\citenamefont {Zöttl}\ and\ \citenamefont {Stark}(2016)}]{Zottl2016}%
  \BibitemOpen
  \bibfield  {author} {\bibinfo {author} {\bibfnamefont {A.}~\bibnamefont {Zöttl}}\ and\ \bibinfo {author} {\bibfnamefont {H.}~\bibnamefont {Stark}},\ }\href@noop {} {\bibfield  {journal} {\bibinfo  {journal} {J. Phys.: Condens. Matter}\ }\textbf {\bibinfo {volume} {28}},\ \bibinfo {pages} {253001} (\bibinfo {year} {2016})}\BibitemShut {NoStop}%
\bibitem [{\citenamefont {Katuri}\ \emph {et~al.}(2017)\citenamefont {Katuri}, \citenamefont {Ma}, \citenamefont {Stanton},\ and\ \citenamefont {S\'anchez}}]{Katuri2017}%
  \BibitemOpen
  \bibfield  {author} {\bibinfo {author} {\bibfnamefont {J.}~\bibnamefont {Katuri}}, \bibinfo {author} {\bibfnamefont {X.}~\bibnamefont {Ma}}, \bibinfo {author} {\bibfnamefont {M.~M.}\ \bibnamefont {Stanton}}, \ and\ \bibinfo {author} {\bibfnamefont {S.}~\bibnamefont {S\'anchez}},\ }\href {\doibase 10.1021/acs.accounts.6b00386} {\bibfield  {journal} {\bibinfo  {journal} {Acc. Chem. Res.}\ }\textbf {\bibinfo {volume} {50}},\ \bibinfo {pages} {2} (\bibinfo {year} {2017})}\BibitemShut {NoStop}%
\bibitem [{\citenamefont {Patra}\ \emph {et~al.}(2013)\citenamefont {Patra}, \citenamefont {Sengupta}, \citenamefont {Duan}, \citenamefont {Zhang}, \citenamefont {Pavlick},\ and\ \citenamefont {Sen}}]{Patra2013}%
  \BibitemOpen
  \bibfield  {author} {\bibinfo {author} {\bibfnamefont {D.}~\bibnamefont {Patra}}, \bibinfo {author} {\bibfnamefont {S.}~\bibnamefont {Sengupta}}, \bibinfo {author} {\bibfnamefont {W.}~\bibnamefont {Duan}}, \bibinfo {author} {\bibfnamefont {H.}~\bibnamefont {Zhang}}, \bibinfo {author} {\bibfnamefont {R.}~\bibnamefont {Pavlick}}, \ and\ \bibinfo {author} {\bibfnamefont {A.}~\bibnamefont {Sen}},\ }\href {\doibase 10.1039/C2NR32600K} {\bibfield  {journal} {\bibinfo  {journal} {Nanoscale}\ }\textbf {\bibinfo {volume} {5}},\ \bibinfo {pages} {1273} (\bibinfo {year} {2013})}\BibitemShut {NoStop}%
\bibitem [{\citenamefont {Han}\ \emph {et~al.}(2018)\citenamefont {Han}, \citenamefont {IV},\ and\ \citenamefont {Velev}}]{Han2018}%
  \BibitemOpen
  \bibfield  {author} {\bibinfo {author} {\bibfnamefont {K.}~\bibnamefont {Han}}, \bibinfo {author} {\bibfnamefont {C.~W.~S.}\ \bibnamefont {IV}}, \ and\ \bibinfo {author} {\bibfnamefont {O.~D.}\ \bibnamefont {Velev}},\ }\href@noop {} {\bibfield  {journal} {\bibinfo  {journal} {Adv. Funct. Mater.}\ }\textbf {\bibinfo {volume} {28}},\ \bibinfo {pages} {1705953} (\bibinfo {year} {2018})}\BibitemShut {NoStop}%
\bibitem [{\citenamefont {Garc\'ia}\ \emph {et~al.}(2013)\citenamefont {Garc\'ia}, \citenamefont {Orozco}, \citenamefont {Guix}, \citenamefont {Gao}, \citenamefont {Sattayasamitsathit}, \citenamefont {Escarpa}, \citenamefont {Merkocic},\ and\ \citenamefont {Wang}}]{Garcia2013}%
  \BibitemOpen
  \bibfield  {author} {\bibinfo {author} {\bibfnamefont {M.}~\bibnamefont {Garc\'ia}}, \bibinfo {author} {\bibfnamefont {J.}~\bibnamefont {Orozco}}, \bibinfo {author} {\bibfnamefont {M.}~\bibnamefont {Guix}}, \bibinfo {author} {\bibfnamefont {W.}~\bibnamefont {Gao}}, \bibinfo {author} {\bibfnamefont {S.}~\bibnamefont {Sattayasamitsathit}}, \bibinfo {author} {\bibfnamefont {A.}~\bibnamefont {Escarpa}}, \bibinfo {author} {\bibfnamefont {A.}~\bibnamefont {Merkocic}}, \ and\ \bibinfo {author} {\bibfnamefont {J.}~\bibnamefont {Wang}},\ }\href {\doibase 10.1039/C2NR32400H} {\bibfield  {journal} {\bibinfo  {journal} {Current Opinion in Colloid and Interface Science}\ }\textbf {\bibinfo {volume} {5}},\ \bibinfo {pages} {1325} (\bibinfo {year} {2013})}\BibitemShut {NoStop}%
\bibitem [{\citenamefont {Restrepo-P\'erez}\ \emph {et~al.}(2014)\citenamefont {Restrepo-P\'erez}, \citenamefont {Soler}, \citenamefont {Mart\'inez-Cisneros}, \citenamefont {S\'anchez},\ and\ \citenamefont {Schmidt}}]{Restrepo2014}%
  \BibitemOpen
  \bibfield  {author} {\bibinfo {author} {\bibfnamefont {L.}~\bibnamefont {Restrepo-P\'erez}}, \bibinfo {author} {\bibfnamefont {L.}~\bibnamefont {Soler}}, \bibinfo {author} {\bibfnamefont {C.}~\bibnamefont {Mart\'inez-Cisneros}}, \bibinfo {author} {\bibfnamefont {S.}~\bibnamefont {S\'anchez}}, \ and\ \bibinfo {author} {\bibfnamefont {O.~G.}\ \bibnamefont {Schmidt}},\ }\href {\doibase 10.1039/c4lc00439f} {\bibfield  {journal} {\bibinfo  {journal} {Lab on a chip}\ }\textbf {\bibinfo {volume} {14}},\ \bibinfo {pages} {2914} (\bibinfo {year} {2014})}\BibitemShut {NoStop}%
\bibitem [{\citenamefont {Gao}\ \emph {et~al.}(2015)\citenamefont {Gao}, \citenamefont {Dong}, \citenamefont {Thamphiwatana}, \citenamefont {Li}, \citenamefont {Gao}, \citenamefont {Zhang},\ and\ \citenamefont {Wang}}]{Gao2015}%
  \BibitemOpen
  \bibfield  {author} {\bibinfo {author} {\bibfnamefont {W.}~\bibnamefont {Gao}}, \bibinfo {author} {\bibfnamefont {R.}~\bibnamefont {Dong}}, \bibinfo {author} {\bibfnamefont {S.}~\bibnamefont {Thamphiwatana}}, \bibinfo {author} {\bibfnamefont {J.}~\bibnamefont {Li}}, \bibinfo {author} {\bibfnamefont {W.}~\bibnamefont {Gao}}, \bibinfo {author} {\bibfnamefont {L.}~\bibnamefont {Zhang}}, \ and\ \bibinfo {author} {\bibfnamefont {J.}~\bibnamefont {Wang}},\ }\href@noop {} {\bibfield  {journal} {\bibinfo  {journal} {ACS Nano}\ }\textbf {\bibinfo {volume} {9}},\ \bibinfo {pages} {117} (\bibinfo {year} {2015})}\BibitemShut {NoStop}%
\bibitem [{\citenamefont {Wang}\ \emph {et~al.}(2018)\citenamefont {Wang}, \citenamefont {Xiong}, \citenamefont {Zheng}, \citenamefont {Zhan},\ and\ \citenamefont {Tang}}]{Wang2018}%
  \BibitemOpen
  \bibfield  {author} {\bibinfo {author} {\bibfnamefont {J.}~\bibnamefont {Wang}}, \bibinfo {author} {\bibfnamefont {Z.}~\bibnamefont {Xiong}}, \bibinfo {author} {\bibfnamefont {J.}~\bibnamefont {Zheng}}, \bibinfo {author} {\bibfnamefont {X.}~\bibnamefont {Zhan}}, \ and\ \bibinfo {author} {\bibfnamefont {J.}~\bibnamefont {Tang}},\ }\href@noop {} {\bibfield  {journal} {\bibinfo  {journal} {Acc. Chem. Res.}\ }\textbf {\bibinfo {volume} {51}},\ \bibinfo {pages} {1957} (\bibinfo {year} {2018})}\BibitemShut {NoStop}%
\bibitem [{\citenamefont {Ceylan}\ \emph {et~al.}(2019)\citenamefont {Ceylan}, \citenamefont {Yasa}, \citenamefont {Yasa}, \citenamefont {Tabak}, \citenamefont {Giltinan},\ and\ \citenamefont {Sitti}}]{Ceylan2019}%
  \BibitemOpen
  \bibfield  {author} {\bibinfo {author} {\bibfnamefont {H.}~\bibnamefont {Ceylan}}, \bibinfo {author} {\bibfnamefont {I.~C.}\ \bibnamefont {Yasa}}, \bibinfo {author} {\bibfnamefont {O.}~\bibnamefont {Yasa}}, \bibinfo {author} {\bibfnamefont {A.~F.}\ \bibnamefont {Tabak}}, \bibinfo {author} {\bibfnamefont {J.}~\bibnamefont {Giltinan}}, \ and\ \bibinfo {author} {\bibfnamefont {M.}~\bibnamefont {Sitti}},\ }\href@noop {} {\bibfield  {journal} {\bibinfo  {journal} {ACS Nano}\ }\textbf {\bibinfo {volume} {13}},\ \bibinfo {pages} {3353} (\bibinfo {year} {2019})}\BibitemShut {NoStop}%
\bibitem [{\citenamefont {Gao}\ \emph {et~al.}(2013)\citenamefont {Gao}, \citenamefont {Feng}, \citenamefont {Pei}, \citenamefont {Gu}, \citenamefont {Li},\ and\ \citenamefont {Wang}}]{Gao2013}%
  \BibitemOpen
  \bibfield  {author} {\bibinfo {author} {\bibfnamefont {W.}~\bibnamefont {Gao}}, \bibinfo {author} {\bibfnamefont {X.}~\bibnamefont {Feng}}, \bibinfo {author} {\bibfnamefont {A.}~\bibnamefont {Pei}}, \bibinfo {author} {\bibfnamefont {Y.}~\bibnamefont {Gu}}, \bibinfo {author} {\bibfnamefont {J.}~\bibnamefont {Li}}, \ and\ \bibinfo {author} {\bibfnamefont {J.}~\bibnamefont {Wang}},\ }\href {\doibase 10.1039/C3NR01458D} {\bibfield  {journal} {\bibinfo  {journal} {Nanoscale}\ }\textbf {\bibinfo {volume} {5}},\ \bibinfo {pages} {4696} (\bibinfo {year} {2013})}\BibitemShut {NoStop}%
\bibitem [{\citenamefont {Wang}\ \emph {et~al.}(2021)\citenamefont {Wang}, \citenamefont {Ji}, \citenamefont {Wang},\ and\ \citenamefont {Zhang}}]{Wang2021}%
  \BibitemOpen
  \bibfield  {author} {\bibinfo {author} {\bibfnamefont {Q.}~\bibnamefont {Wang}}, \bibinfo {author} {\bibfnamefont {F.}~\bibnamefont {Ji}}, \bibinfo {author} {\bibfnamefont {D.~S.}\ \bibnamefont {Wang}}, \ and\ \bibinfo {author} {\bibfnamefont {L.}~\bibnamefont {Zhang}},\ }\href@noop {} {\bibfield  {journal} {\bibinfo  {journal} {ChemNanoMat}\ }\textbf {\bibinfo {volume} {7}},\ \bibinfo {pages} {600} (\bibinfo {year} {2021})}\BibitemShut {NoStop}%
\bibitem [{\citenamefont {Xu}\ \emph {et~al.}(2021)\citenamefont {Xu}, \citenamefont {Yuan},\ and\ \citenamefont {Ma}}]{Xu2021}%
  \BibitemOpen
  \bibfield  {author} {\bibinfo {author} {\bibfnamefont {D.}~\bibnamefont {Xu}}, \bibinfo {author} {\bibfnamefont {H.}~\bibnamefont {Yuan}}, \ and\ \bibinfo {author} {\bibfnamefont {X.}~\bibnamefont {Ma}},\ }\href@noop {} {\bibfield  {journal} {\bibinfo  {journal} {ChemNanoMat}\ }\textbf {\bibinfo {volume} {7}},\ \bibinfo {pages} {439} (\bibinfo {year} {2021})}\BibitemShut {NoStop}%
\bibitem [{\citenamefont {Ebbens}(2016)}]{Ebbens2016}%
  \BibitemOpen
  \bibfield  {author} {\bibinfo {author} {\bibfnamefont {S.~J.}\ \bibnamefont {Ebbens}},\ }\href {\doibase 10.1016/j.cocis.2015.10.003} {\bibfield  {journal} {\bibinfo  {journal} {Current Opinion in Colloid and Interface Science}\ }\textbf {\bibinfo {volume} {21}},\ \bibinfo {pages} {14} (\bibinfo {year} {2016})}\BibitemShut {NoStop}%
\bibitem [{\citenamefont {Bishop}\ \emph {et~al.}(2023)\citenamefont {Bishop}, \citenamefont {Biswal},\ and\ \citenamefont {Bharti}}]{Bishop2023}%
  \BibitemOpen
  \bibfield  {author} {\bibinfo {author} {\bibfnamefont {K.~J.}\ \bibnamefont {Bishop}}, \bibinfo {author} {\bibfnamefont {S.~L.}\ \bibnamefont {Biswal}}, \ and\ \bibinfo {author} {\bibfnamefont {B.}~\bibnamefont {Bharti}},\ }\href@noop {} {\bibfield  {journal} {\bibinfo  {journal} {Annu. Rev. Chem. Biomol. Eng.}\ }\textbf {\bibinfo {volume} {14}},\ \bibinfo {pages} {1} (\bibinfo {year} {2023})}\BibitemShut {NoStop}%
\bibitem [{\citenamefont {Tsang}\ \emph {et~al.}(2020)\citenamefont {Tsang}, \citenamefont {Demir}, \citenamefont {Ding},\ and\ \citenamefont {Pak}}]{Tsang2020}%
  \BibitemOpen
  \bibfield  {author} {\bibinfo {author} {\bibfnamefont {A.~C.~H.}\ \bibnamefont {Tsang}}, \bibinfo {author} {\bibfnamefont {E.}~\bibnamefont {Demir}}, \bibinfo {author} {\bibfnamefont {Y.}~\bibnamefont {Ding}}, \ and\ \bibinfo {author} {\bibfnamefont {O.~S.}\ \bibnamefont {Pak}},\ }\href {\doibase 10.1002/aisy.201900137} {\bibfield  {journal} {\bibinfo  {journal} {Adv. Intell. Syst.}\ }\textbf {\bibinfo {volume} {2}},\ \bibinfo {pages} {1900137} (\bibinfo {year} {2020})}\BibitemShut {NoStop}%
\bibitem [{\citenamefont {Mijalkov}\ and\ \citenamefont {Volpe}(2013)}]{Mijalkov2013}%
  \BibitemOpen
  \bibfield  {author} {\bibinfo {author} {\bibfnamefont {M.}~\bibnamefont {Mijalkov}}\ and\ \bibinfo {author} {\bibfnamefont {G.}~\bibnamefont {Volpe}},\ }\href {\doibase 10.1039/C3SM27923E} {\bibfield  {journal} {\bibinfo  {journal} {Soft Matter}\ }\textbf {\bibinfo {volume} {9}},\ \bibinfo {pages} {6376} (\bibinfo {year} {2013})}\BibitemShut {NoStop}%
\bibitem [{\citenamefont {Katuri}\ \emph {et~al.}(2018)\citenamefont {Katuri}, \citenamefont {Caballero}, \citenamefont {Voituriez}, \citenamefont {Samitier},\ and\ \citenamefont {S\'anchez}}]{Katuri2018}%
  \BibitemOpen
  \bibfield  {author} {\bibinfo {author} {\bibfnamefont {J.}~\bibnamefont {Katuri}}, \bibinfo {author} {\bibfnamefont {D.}~\bibnamefont {Caballero}}, \bibinfo {author} {\bibfnamefont {R.}~\bibnamefont {Voituriez}}, \bibinfo {author} {\bibfnamefont {J.}~\bibnamefont {Samitier}}, \ and\ \bibinfo {author} {\bibfnamefont {S.}~\bibnamefont {S\'anchez}},\ }\href@noop {} {\bibfield  {journal} {\bibinfo  {journal} {ACS nano}\ }\textbf {\bibinfo {volume} {12}},\ \bibinfo {pages} {7282} (\bibinfo {year} {2018})}\BibitemShut {NoStop}%
\bibitem [{\citenamefont {Maggi}\ \emph {et~al.}(2015)\citenamefont {Maggi}, \citenamefont {Simmchen}, \citenamefont {Saglimbeni}, \citenamefont {Katuri}, \citenamefont {Dipalo}, \citenamefont {Angelis}, \citenamefont {Sanchez},\ and\ \citenamefont {Leonardo}}]{Maggi2015}%
  \BibitemOpen
  \bibfield  {author} {\bibinfo {author} {\bibfnamefont {C.}~\bibnamefont {Maggi}}, \bibinfo {author} {\bibfnamefont {J.}~\bibnamefont {Simmchen}}, \bibinfo {author} {\bibfnamefont {F.}~\bibnamefont {Saglimbeni}}, \bibinfo {author} {\bibfnamefont {J.}~\bibnamefont {Katuri}}, \bibinfo {author} {\bibfnamefont {M.}~\bibnamefont {Dipalo}}, \bibinfo {author} {\bibfnamefont {F.~D.}\ \bibnamefont {Angelis}}, \bibinfo {author} {\bibfnamefont {S.}~\bibnamefont {Sanchez}}, \ and\ \bibinfo {author} {\bibfnamefont {R.~D.}\ \bibnamefont {Leonardo}},\ }\href@noop {} {\bibfield  {journal} {\bibinfo  {journal} {Small}\ }\textbf {\bibinfo {volume} {12}},\ \bibinfo {pages} {446} (\bibinfo {year} {2015})}\BibitemShut {NoStop}%
\bibitem [{\citenamefont {Reichhardt}\ and\ \citenamefont {Reichhardt}(2017)}]{Reichhardt2017}%
  \BibitemOpen
  \bibfield  {author} {\bibinfo {author} {\bibfnamefont {C.~O.}\ \bibnamefont {Reichhardt}}\ and\ \bibinfo {author} {\bibfnamefont {C.}~\bibnamefont {Reichhardt}},\ }\href@noop {} {\bibfield  {journal} {\bibinfo  {journal} {Annu. Rev. Condens. Matter Phys.}\ }\textbf {\bibinfo {volume} {8}},\ \bibinfo {pages} {51} (\bibinfo {year} {2017})}\BibitemShut {NoStop}%
\bibitem [{\citenamefont {Kaiser}\ \emph {et~al.}(2012)\citenamefont {Kaiser}, \citenamefont {Wensink},\ and\ \citenamefont {Löwen}}]{Kaiser2012}%
  \BibitemOpen
  \bibfield  {author} {\bibinfo {author} {\bibfnamefont {A.}~\bibnamefont {Kaiser}}, \bibinfo {author} {\bibfnamefont {H.~H.}\ \bibnamefont {Wensink}}, \ and\ \bibinfo {author} {\bibfnamefont {H.}~\bibnamefont {Löwen}},\ }\href@noop {} {\bibfield  {journal} {\bibinfo  {journal} {Phys. Rev. Lett.}\ }\textbf {\bibinfo {volume} {108}},\ \bibinfo {pages} {268307} (\bibinfo {year} {2012})}\BibitemShut {NoStop}%
\bibitem [{\citenamefont {Palacios}\ \emph {et~al.}(2021)\citenamefont {Palacios}, \citenamefont {Tchoumakov}, \citenamefont {Guix}, \citenamefont {Pagonabarraga}, \citenamefont {S\'anchez},\ and\ \citenamefont {Grushin}}]{Palacios2021}%
  \BibitemOpen
  \bibfield  {author} {\bibinfo {author} {\bibfnamefont {L.~S.}\ \bibnamefont {Palacios}}, \bibinfo {author} {\bibfnamefont {S.}~\bibnamefont {Tchoumakov}}, \bibinfo {author} {\bibfnamefont {M.}~\bibnamefont {Guix}}, \bibinfo {author} {\bibfnamefont {I.}~\bibnamefont {Pagonabarraga}}, \bibinfo {author} {\bibfnamefont {S.}~\bibnamefont {S\'anchez}}, \ and\ \bibinfo {author} {\bibfnamefont {A.~G.}\ \bibnamefont {Grushin}},\ }\href@noop {} {\bibfield  {journal} {\bibinfo  {journal} {Nat Commun}\ }\textbf {\bibinfo {volume} {12}},\ \bibinfo {pages} {4691} (\bibinfo {year} {2021})}\BibitemShut {NoStop}%
\bibitem [{\citenamefont {Ebbens}\ and\ \citenamefont {Howse}(2011)}]{Ebbens2011}%
  \BibitemOpen
  \bibfield  {author} {\bibinfo {author} {\bibfnamefont {S.~J.}\ \bibnamefont {Ebbens}}\ and\ \bibinfo {author} {\bibfnamefont {J.~R.}\ \bibnamefont {Howse}},\ }\href {\doibase 10.1021/la2033127} {\bibfield  {journal} {\bibinfo  {journal} {Langmuir}\ }\textbf {\bibinfo {volume} {27}},\ \bibinfo {pages} {12293} (\bibinfo {year} {2011})}\BibitemShut {NoStop}%
\bibitem [{\citenamefont {Campbell}\ and\ \citenamefont {Ebbens}(2013)}]{Campbell2013}%
  \BibitemOpen
  \bibfield  {author} {\bibinfo {author} {\bibfnamefont {A.~I.}\ \bibnamefont {Campbell}}\ and\ \bibinfo {author} {\bibfnamefont {S.~J.}\ \bibnamefont {Ebbens}},\ }\href@noop {} {\bibfield  {journal} {\bibinfo  {journal} {Langmuir}\ }\textbf {\bibinfo {volume} {29}},\ \bibinfo {pages} {14066} (\bibinfo {year} {2013})}\BibitemShut {NoStop}%
\bibitem [{\citenamefont {Ketzetzi}\ \emph {et~al.}(2020)\citenamefont {Ketzetzi}, \citenamefont {de~Graaf},\ and\ \citenamefont {Kraft}}]{Ketzetzi2020sep}%
  \BibitemOpen
  \bibfield  {author} {\bibinfo {author} {\bibfnamefont {S.}~\bibnamefont {Ketzetzi}}, \bibinfo {author} {\bibfnamefont {J.}~\bibnamefont {de~Graaf}}, \ and\ \bibinfo {author} {\bibfnamefont {D.~J.}\ \bibnamefont {Kraft}},\ }\href {\doibase 10.1103/PhysRevLett.125.238001} {\bibfield  {journal} {\bibinfo  {journal} {Phys. Rev. Lett.}\ }\textbf {\bibinfo {volume} {125}},\ \bibinfo {pages} {238001} (\bibinfo {year} {2020})}\BibitemShut {NoStop}%
\bibitem [{\citenamefont {Bailey}\ \emph {et~al.}(2024)\citenamefont {Bailey}, \citenamefont {Gutiérrez}, \citenamefont {Martín-Roca}, \citenamefont {Niggel}, \citenamefont {Carrasco-Fadanelli}, \citenamefont {Buttinoni}, \citenamefont {Pagonabarraga}, \citenamefont {Isa},\ and\ \citenamefont {Valeriani}}]{Bailey2024}%
  \BibitemOpen
  \bibfield  {author} {\bibinfo {author} {\bibfnamefont {M.~R.}\ \bibnamefont {Bailey}}, \bibinfo {author} {\bibfnamefont {C.~M.~B.}\ \bibnamefont {Gutiérrez}}, \bibinfo {author} {\bibfnamefont {J.}~\bibnamefont {Martín-Roca}}, \bibinfo {author} {\bibfnamefont {V.}~\bibnamefont {Niggel}}, \bibinfo {author} {\bibfnamefont {V.}~\bibnamefont {Carrasco-Fadanelli}}, \bibinfo {author} {\bibfnamefont {I.}~\bibnamefont {Buttinoni}}, \bibinfo {author} {\bibfnamefont {I.}~\bibnamefont {Pagonabarraga}}, \bibinfo {author} {\bibfnamefont {L.}~\bibnamefont {Isa}}, \ and\ \bibinfo {author} {\bibfnamefont {C.}~\bibnamefont {Valeriani}},\ }\href@noop {} {\bibfield  {journal} {\bibinfo  {journal} {Nanoscale}\ }\textbf {\bibinfo {volume} {16}},\ \bibinfo {pages} {2444} (\bibinfo {year} {2024})}\BibitemShut {NoStop}%
\bibitem [{\citenamefont {Carrasco-Fadanelli}\ and\ \citenamefont {Buttinoni}(2023)}]{Carrasco2023}%
  \BibitemOpen
  \bibfield  {author} {\bibinfo {author} {\bibfnamefont {V.}~\bibnamefont {Carrasco-Fadanelli}}\ and\ \bibinfo {author} {\bibfnamefont {I.}~\bibnamefont {Buttinoni}},\ }\href@noop {} {\bibfield  {journal} {\bibinfo  {journal} {Phys. Rev. Research}\ }\textbf {\bibinfo {volume} {5}},\ \bibinfo {pages} {L012018} (\bibinfo {year} {2023})}\BibitemShut {NoStop}%
\bibitem [{\citenamefont {Das}\ \emph {et~al.}(2015)\citenamefont {Das}, \citenamefont {Garg}, \citenamefont {Campbell}, \citenamefont {Howse}, \citenamefont {Sen}, \citenamefont {Velegol}, \citenamefont {Golestanian},\ and\ \citenamefont {Ebbens}}]{Das2015}%
  \BibitemOpen
  \bibfield  {author} {\bibinfo {author} {\bibfnamefont {S.}~\bibnamefont {Das}}, \bibinfo {author} {\bibfnamefont {A.}~\bibnamefont {Garg}}, \bibinfo {author} {\bibfnamefont {A.~I.}\ \bibnamefont {Campbell}}, \bibinfo {author} {\bibfnamefont {J.}~\bibnamefont {Howse}}, \bibinfo {author} {\bibfnamefont {A.}~\bibnamefont {Sen}}, \bibinfo {author} {\bibfnamefont {D.}~\bibnamefont {Velegol}}, \bibinfo {author} {\bibfnamefont {R.}~\bibnamefont {Golestanian}}, \ and\ \bibinfo {author} {\bibfnamefont {S.~J.}\ \bibnamefont {Ebbens}},\ }\href {\doibase 10.1038/ncomms9999} {\bibfield  {journal} {\bibinfo  {journal} {Nat. Commun.}\ }\textbf {\bibinfo {volume} {6}},\ \bibinfo {pages} {8999} (\bibinfo {year} {2015})}\BibitemShut {NoStop}%
\bibitem [{\citenamefont {Simmchen}\ \emph {et~al.}(2016)\citenamefont {Simmchen}, \citenamefont {Katuri}, \citenamefont {Uspal}, \citenamefont {Popescu}, \citenamefont {Tasinkevych},\ and\ \citenamefont {S\'anchez}}]{Simmchen2016}%
  \BibitemOpen
  \bibfield  {author} {\bibinfo {author} {\bibfnamefont {J.}~\bibnamefont {Simmchen}}, \bibinfo {author} {\bibfnamefont {J.}~\bibnamefont {Katuri}}, \bibinfo {author} {\bibfnamefont {W.~E.}\ \bibnamefont {Uspal}}, \bibinfo {author} {\bibfnamefont {M.~N.}\ \bibnamefont {Popescu}}, \bibinfo {author} {\bibfnamefont {M.}~\bibnamefont {Tasinkevych}}, \ and\ \bibinfo {author} {\bibfnamefont {S.}~\bibnamefont {S\'anchez}},\ }\href@noop {} {\bibfield  {journal} {\bibinfo  {journal} {Nat. Commun.}\ }\textbf {\bibinfo {volume} {7}},\ \bibinfo {pages} {10598} (\bibinfo {year} {2016})}\BibitemShut {NoStop}%
\bibitem [{\citenamefont {Yu}\ \emph {et~al.}(2016)\citenamefont {Yu}, \citenamefont {Kopach}, \citenamefont {Misko}, \citenamefont {Vasylenko}, \citenamefont {Makarov}, \citenamefont {Marchesoni}, \citenamefont {Nori}, \citenamefont {Baraban},\ and\ \citenamefont {Cuniberti}}]{Yu2016}%
  \BibitemOpen
  \bibfield  {author} {\bibinfo {author} {\bibfnamefont {H.}~\bibnamefont {Yu}}, \bibinfo {author} {\bibfnamefont {A.}~\bibnamefont {Kopach}}, \bibinfo {author} {\bibfnamefont {V.~R.}\ \bibnamefont {Misko}}, \bibinfo {author} {\bibfnamefont {A.~A.}\ \bibnamefont {Vasylenko}}, \bibinfo {author} {\bibfnamefont {D.}~\bibnamefont {Makarov}}, \bibinfo {author} {\bibfnamefont {F.}~\bibnamefont {Marchesoni}}, \bibinfo {author} {\bibfnamefont {F.}~\bibnamefont {Nori}}, \bibinfo {author} {\bibfnamefont {L.}~\bibnamefont {Baraban}}, \ and\ \bibinfo {author} {\bibfnamefont {G.}~\bibnamefont {Cuniberti}},\ }\href {\doibase 10.1002/smll.201602039} {\bibfield  {journal} {\bibinfo  {journal} {Small}\ }\textbf {\bibinfo {volume} {12}},\ \bibinfo {pages} {5882} (\bibinfo {year} {2016})}\BibitemShut {NoStop}%
\bibitem [{\citenamefont {Liu}\ \emph {et~al.}(2016)\citenamefont {Liu}, \citenamefont {Zhou}, \citenamefont {Wang},\ and\ \citenamefont {Zhang}}]{Liu2016}%
  \BibitemOpen
  \bibfield  {author} {\bibinfo {author} {\bibfnamefont {C.}~\bibnamefont {Liu}}, \bibinfo {author} {\bibfnamefont {C.}~\bibnamefont {Zhou}}, \bibinfo {author} {\bibfnamefont {W.}~\bibnamefont {Wang}}, \ and\ \bibinfo {author} {\bibfnamefont {H.~P.}\ \bibnamefont {Zhang}},\ }\href {\doibase 10.1103/PhysRevLett.117.198001} {\bibfield  {journal} {\bibinfo  {journal} {Phys. Rev. Lett.}\ }\textbf {\bibinfo {volume} {117}},\ \bibinfo {pages} {198001} (\bibinfo {year} {2016})}\BibitemShut {NoStop}%
\bibitem [{\citenamefont {Takagi}\ \emph {et~al.}(2014)\citenamefont {Takagi}, \citenamefont {Palacci}, \citenamefont {Braunschweig}, \citenamefont {Shelley},\ and\ \citenamefont {Zhang}}]{Takagi2014}%
  \BibitemOpen
  \bibfield  {author} {\bibinfo {author} {\bibfnamefont {D.}~\bibnamefont {Takagi}}, \bibinfo {author} {\bibfnamefont {J.}~\bibnamefont {Palacci}}, \bibinfo {author} {\bibfnamefont {A.~B.}\ \bibnamefont {Braunschweig}}, \bibinfo {author} {\bibfnamefont {M.~J.}\ \bibnamefont {Shelley}}, \ and\ \bibinfo {author} {\bibfnamefont {J.}~\bibnamefont {Zhang}},\ }\href {\doibase 10.1039/c3sm52815d} {\bibfield  {journal} {\bibinfo  {journal} {Soft Matter}\ }\textbf {\bibinfo {volume} {10}},\ \bibinfo {pages} {1784} (\bibinfo {year} {2014})}\BibitemShut {NoStop}%
\bibitem [{\citenamefont {Brown}\ \emph {et~al.}(2016)\citenamefont {Brown}, \citenamefont {Vladescu}, \citenamefont {Dawson}, \citenamefont {Vissers}, \citenamefont {Schwarz-Linek}, \citenamefont {Lintuvuori},\ and\ \citenamefont {Poon}}]{Brown2016}%
  \BibitemOpen
  \bibfield  {author} {\bibinfo {author} {\bibfnamefont {A.~T.}\ \bibnamefont {Brown}}, \bibinfo {author} {\bibfnamefont {I.~D.}\ \bibnamefont {Vladescu}}, \bibinfo {author} {\bibfnamefont {A.}~\bibnamefont {Dawson}}, \bibinfo {author} {\bibfnamefont {T.}~\bibnamefont {Vissers}}, \bibinfo {author} {\bibfnamefont {J.}~\bibnamefont {Schwarz-Linek}}, \bibinfo {author} {\bibfnamefont {J.~S.}\ \bibnamefont {Lintuvuori}}, \ and\ \bibinfo {author} {\bibfnamefont {W.~C.~K.}\ \bibnamefont {Poon}},\ }\href {\doibase 10.1039/c5sm01831e} {\bibfield  {journal} {\bibinfo  {journal} {Soft Matter}\ }\textbf {\bibinfo {volume} {12}},\ \bibinfo {pages} {131} (\bibinfo {year} {2016})}\BibitemShut {NoStop}%
\bibitem [{\citenamefont {van Baalen}\ \emph {et~al.}(2023)\citenamefont {van Baalen}, \citenamefont {Uspal}, \citenamefont {Popescu},\ and\ \citenamefont {Isa}}]{Baalen2023}%
  \BibitemOpen
  \bibfield  {author} {\bibinfo {author} {\bibfnamefont {C.}~\bibnamefont {van Baalen}}, \bibinfo {author} {\bibfnamefont {W.~E.}\ \bibnamefont {Uspal}}, \bibinfo {author} {\bibfnamefont {M.~N.}\ \bibnamefont {Popescu}}, \ and\ \bibinfo {author} {\bibfnamefont {L.}~\bibnamefont {Isa}},\ }\href@noop {} {\bibfield  {journal} {\bibinfo  {journal} {Soft Matter}\ }\textbf {\bibinfo {volume} {19}},\ \bibinfo {pages} {8790} (\bibinfo {year} {2023})}\BibitemShut {NoStop}%
\bibitem [{\citenamefont {Ketzetzi}\ \emph {et~al.}(2022)\citenamefont {Ketzetzi}, \citenamefont {Rinaldin}, \citenamefont {Dröge}, \citenamefont {de~Graaf},\ and\ \citenamefont {Kraft}}]{Ketzetzi2022}%
  \BibitemOpen
  \bibfield  {author} {\bibinfo {author} {\bibfnamefont {S.}~\bibnamefont {Ketzetzi}}, \bibinfo {author} {\bibfnamefont {M.}~\bibnamefont {Rinaldin}}, \bibinfo {author} {\bibfnamefont {P.}~\bibnamefont {Dröge}}, \bibinfo {author} {\bibfnamefont {J.}~\bibnamefont {de~Graaf}}, \ and\ \bibinfo {author} {\bibfnamefont {D.~J.}\ \bibnamefont {Kraft}},\ }\href {\doibase 10.1038/s41467-022-29430-1} {\bibfield  {journal} {\bibinfo  {journal} {Nat Commun}\ }\textbf {\bibinfo {volume} {13}},\ \bibinfo {pages} {1772} (\bibinfo {year} {2022})}\BibitemShut {NoStop}%
\bibitem [{\citenamefont {Wykes}\ \emph {et~al.}(2017)\citenamefont {Wykes}, \citenamefont {Zhong}, \citenamefont {Tong}, \citenamefont {Adachi}, \citenamefont {Liu}, \citenamefont {Ristroph}, \citenamefont {Ward}, \citenamefont {Shelley},\ and\ \citenamefont {Zhang}}]{Wykes2017}%
  \BibitemOpen
  \bibfield  {author} {\bibinfo {author} {\bibfnamefont {M.~S.~D.}\ \bibnamefont {Wykes}}, \bibinfo {author} {\bibfnamefont {X.}~\bibnamefont {Zhong}}, \bibinfo {author} {\bibfnamefont {J.}~\bibnamefont {Tong}}, \bibinfo {author} {\bibfnamefont {T.}~\bibnamefont {Adachi}}, \bibinfo {author} {\bibfnamefont {Y.}~\bibnamefont {Liu}}, \bibinfo {author} {\bibfnamefont {L.}~\bibnamefont {Ristroph}}, \bibinfo {author} {\bibfnamefont {M.~D.}\ \bibnamefont {Ward}}, \bibinfo {author} {\bibfnamefont {M.~J.}\ \bibnamefont {Shelley}}, \ and\ \bibinfo {author} {\bibfnamefont {J.}~\bibnamefont {Zhang}},\ }\href {\doibase 10.1039/C7SM00203C} {\bibfield  {journal} {\bibinfo  {journal} {Soft Matter}\ }\textbf {\bibinfo {volume} {13}},\ \bibinfo {pages} {4681} (\bibinfo {year} {2017})}\BibitemShut {NoStop}%
\bibitem [{\citenamefont {Tanuku}\ \emph {et~al.}(2023)\citenamefont {Tanuku}, \citenamefont {Vogel}, \citenamefont {Palberg},\ and\ \citenamefont {Buttinoni}}]{Tanuku2023}%
  \BibitemOpen
  \bibfield  {author} {\bibinfo {author} {\bibfnamefont {V.~M. S.~G.}\ \bibnamefont {Tanuku}}, \bibinfo {author} {\bibfnamefont {P.}~\bibnamefont {Vogel}}, \bibinfo {author} {\bibfnamefont {T.}~\bibnamefont {Palberg}}, \ and\ \bibinfo {author} {\bibfnamefont {I.}~\bibnamefont {Buttinoni}},\ }\href@noop {} {\bibfield  {journal} {\bibinfo  {journal} {Soft Matter}\ }\textbf {\bibinfo {volume} {19}},\ \bibinfo {pages} {5452} (\bibinfo {year} {2023})}\BibitemShut {NoStop}%
\bibitem [{\citenamefont {Trivedi}\ \emph {et~al.}(2022)\citenamefont {Trivedi}, \citenamefont {Saxena}, \citenamefont {Ng}, \citenamefont {Sapienza},\ and\ \citenamefont {Volpe}}]{Trivedi2022}%
  \BibitemOpen
  \bibfield  {author} {\bibinfo {author} {\bibfnamefont {M.}~\bibnamefont {Trivedi}}, \bibinfo {author} {\bibfnamefont {D.}~\bibnamefont {Saxena}}, \bibinfo {author} {\bibfnamefont {W.~K.}\ \bibnamefont {Ng}}, \bibinfo {author} {\bibfnamefont {R.}~\bibnamefont {Sapienza}}, \ and\ \bibinfo {author} {\bibfnamefont {G.}~\bibnamefont {Volpe}},\ }\href@noop {} {\bibfield  {journal} {\bibinfo  {journal} {Nature Physics}\ }\textbf {\bibinfo {volume} {18}},\ \bibinfo {pages} {939} (\bibinfo {year} {2022})}\BibitemShut {NoStop}%
\bibitem [{\citenamefont {IV}\ and\ \citenamefont {Velev}(2017)}]{Shields2017}%
  \BibitemOpen
  \bibfield  {author} {\bibinfo {author} {\bibfnamefont {C.~W.~S.}\ \bibnamefont {IV}}\ and\ \bibinfo {author} {\bibfnamefont {O.~D.}\ \bibnamefont {Velev}},\ }\href@noop {} {\bibfield  {journal} {\bibinfo  {journal} {Chem}\ }\textbf {\bibinfo {volume} {3}},\ \bibinfo {pages} {539} (\bibinfo {year} {2017})}\BibitemShut {NoStop}%
\bibitem [{\citenamefont {Squires}\ and\ \citenamefont {Bazant}(2006)}]{Squires2006}%
  \BibitemOpen
  \bibfield  {author} {\bibinfo {author} {\bibfnamefont {T.~M.}\ \bibnamefont {Squires}}\ and\ \bibinfo {author} {\bibfnamefont {M.~Z.}\ \bibnamefont {Bazant}},\ }\href@noop {} {\bibfield  {journal} {\bibinfo  {journal} {J. Fluid Mech.}\ }\textbf {\bibinfo {volume} {560}},\ \bibinfo {pages} {65} (\bibinfo {year} {2006})}\BibitemShut {NoStop}%
\bibitem [{\citenamefont {Ristenpart}\ \emph {et~al.}(2007)\citenamefont {Ristenpart}, \citenamefont {Aksay},\ and\ \citenamefont {Saville}}]{Ristenpart2007}%
  \BibitemOpen
  \bibfield  {author} {\bibinfo {author} {\bibfnamefont {W.}~\bibnamefont {Ristenpart}}, \bibinfo {author} {\bibfnamefont {I.~A.}\ \bibnamefont {Aksay}}, \ and\ \bibinfo {author} {\bibfnamefont {D.}~\bibnamefont {Saville}},\ }\href@noop {} {\bibfield  {journal} {\bibinfo  {journal} {Journal of Fluid Mechanics}\ }\textbf {\bibinfo {volume} {575}},\ \bibinfo {pages} {83} (\bibinfo {year} {2007})}\BibitemShut {NoStop}%
\bibitem [{\citenamefont {Ma}\ \emph {et~al.}(2015)\citenamefont {Ma}, \citenamefont {Yang}, \citenamefont {Zhao2},\ and\ \citenamefont {Wu}}]{Ma2015}%
  \BibitemOpen
  \bibfield  {author} {\bibinfo {author} {\bibfnamefont {F.}~\bibnamefont {Ma}}, \bibinfo {author} {\bibfnamefont {X.}~\bibnamefont {Yang}}, \bibinfo {author} {\bibfnamefont {H.}~\bibnamefont {Zhao2}}, \ and\ \bibinfo {author} {\bibfnamefont {N.}~\bibnamefont {Wu}},\ }\href@noop {} {\bibfield  {journal} {\bibinfo  {journal} {Phys. Rev. Lett.}\ }\textbf {\bibinfo {volume} {115}},\ \bibinfo {pages} {208302} (\bibinfo {year} {2015})}\BibitemShut {NoStop}%
\bibitem [{\citenamefont {Yang}\ \emph {et~al.}(2019)\citenamefont {Yang}, \citenamefont {Johnson},\ and\ \citenamefont {Wu}}]{Yang2019}%
  \BibitemOpen
  \bibfield  {author} {\bibinfo {author} {\bibfnamefont {X.}~\bibnamefont {Yang}}, \bibinfo {author} {\bibfnamefont {S.}~\bibnamefont {Johnson}}, \ and\ \bibinfo {author} {\bibfnamefont {N.}~\bibnamefont {Wu}},\ }\href@noop {} {\bibfield  {journal} {\bibinfo  {journal} {Adv. Intell. Syst.}\ }\textbf {\bibinfo {volume} {1}},\ \bibinfo {pages} {1900096} (\bibinfo {year} {2019})}\BibitemShut {NoStop}%
\bibitem [{\citenamefont {Ristenpart}\ \emph {et~al.}(2004)\citenamefont {Ristenpart}, \citenamefont {Aksay},\ and\ \citenamefont {Saville}}]{Ristenpart2004}%
  \BibitemOpen
  \bibfield  {author} {\bibinfo {author} {\bibfnamefont {W.}~\bibnamefont {Ristenpart}}, \bibinfo {author} {\bibfnamefont {I.~A.}\ \bibnamefont {Aksay}}, \ and\ \bibinfo {author} {\bibfnamefont {D.}~\bibnamefont {Saville}},\ }\href@noop {} {\bibfield  {journal} {\bibinfo  {journal} {Physical Review E}\ }\textbf {\bibinfo {volume} {69}},\ \bibinfo {pages} {021405} (\bibinfo {year} {2004})}\BibitemShut {NoStop}%
\bibitem [{\citenamefont {Crocker}\ and\ \citenamefont {Grier}(1996)}]{Crocker1996}%
  \BibitemOpen
  \bibfield  {author} {\bibinfo {author} {\bibfnamefont {J.~C.}\ \bibnamefont {Crocker}}\ and\ \bibinfo {author} {\bibfnamefont {D.~G.}\ \bibnamefont {Grier}},\ }\href@noop {} {\bibfield  {journal} {\bibinfo  {journal} {Journal of Colloid and Interface Science}\ }\textbf {\bibinfo {volume} {179}},\ \bibinfo {pages} {298} (\bibinfo {year} {1996})}\BibitemShut {NoStop}%
\bibitem [{\citenamefont {Allan}\ \emph {et~al.}(2018)\citenamefont {Allan}, \citenamefont {Caswell}, \citenamefont {Keim},\ and\ \citenamefont {van~der Wel}}]{Trackpy}%
  \BibitemOpen
  \bibfield  {author} {\bibinfo {author} {\bibfnamefont {D.~B.}\ \bibnamefont {Allan}}, \bibinfo {author} {\bibfnamefont {T.}~\bibnamefont {Caswell}}, \bibinfo {author} {\bibfnamefont {N.~C.}\ \bibnamefont {Keim}}, \ and\ \bibinfo {author} {\bibfnamefont {C.~M.}\ \bibnamefont {van~der Wel}},\ }\href {\doibase 10.5281/zenodo.1226458} {\enquote {\bibinfo {title} {trackpy: Trackpy v0.4.1},}\ } (\bibinfo {year} {2018})\BibitemShut {NoStop}%
\end{thebibliography}%

\end{document}